\documentclass{aa}

\usepackage{txfonts}
\usepackage{graphicx}
\usepackage{natbib}
\bibpunct{(}{)}{;}{a}{}{,}

\newcommand{\mc}[1]{\multicolumn{1}{c}{#1}}

\begin{document}

\title{New limits on the density of the extragalactic background light in the optical to the far infrared from the spectra of all known TeV blazars}

\author{Daniel Mazin\inst{1}
  \and Martin Raue\inst{2}}

\offprints{D. Mazin, \email{mazin@mppmu.mpg.de}, M. Raue, \email{martin.raue@desy.de}}

\institute{Max-Planck-Institut f\"ur Physik, Foehringer Ring 6, 80805 Munich, Germany
  \and Institut f\"ur Experimentalphysik, Universit\"at Hamburg, Luruper Chaussee 149, 22761 Hamburg, Germany}

\date{Received  / Accepted }

\abstract {}
{The extragalactic background light (EBL) in the ultraviolet to far-infrared wavelength region carries important information about galaxy and star formation history. Direct measurements are difficult, especially in the mid infrared region. We derive limits on the EBL density from the energy spectra of distant sources of very high energetic $\gamma$-rays  (VHE $\gamma$-rays).} 
{The VHE $\gamma$-rays are attenuated by the photons of the EBL via pair production, which leaves an imprint on the measured spectra from distant sources. So far, there are 14 detected extragalactic sources of VHE $\gamma$-rays, 13 of which are TeV blazars. With physical assumptions about the intrinsic spectra of these sources, limits on the EBL can be derived. In this paper we present a new method of deriving constraints on the EBL. Here, we use only very basic assumptions about TeV blazar physics and no pre-defined EBL model, but instead a large number of generic shapes constructed from a grid in EBL density vs. wavelength. In our study we utilize spectral data from all known TeV blazars, making this the most complete study so far.} 
{We derive limits on the EBL for three individual TeV blazar spectra (Mkn\,501, H\,1426+428, 1ES\,1101-232) and for all spectra combined. Combining the results from individual spectra leads to significantly stronger constraints over a wide wavelength range from the optical ($\sim1\,\mu$m) to the far-infrared ($\sim 80\,\mu$m). The limits are only a factor of 2 to 3 above the absolute lower limits  derived from source counts. In the mid-infrared our limits are the strongest constraints derived from TeV blazar spectra so far over an extended wavelength range. A high density of the EBL around 1$\,\mu$m, reported by direct detection experiments, can be excluded.}
{Our results can be interpreted in two ways. (i) The EBL is almost resolved by source counts, leaving only a little room for additional components, such as the first stars, or (ii) the assumptions about the underlying physics are not valid, which would require substantial changes in the standard emission models of TeV blazars.}


\authorrunning{D. Mazin and M. Raue}
\titlerunning{New limits on the EBL density from TeV blazars}

\maketitle

\section{Introduction}

\begin{figure*}[t,b]
\centering
\begin{minipage}[c]{0.6\textwidth}
\includegraphics[width=\textwidth]{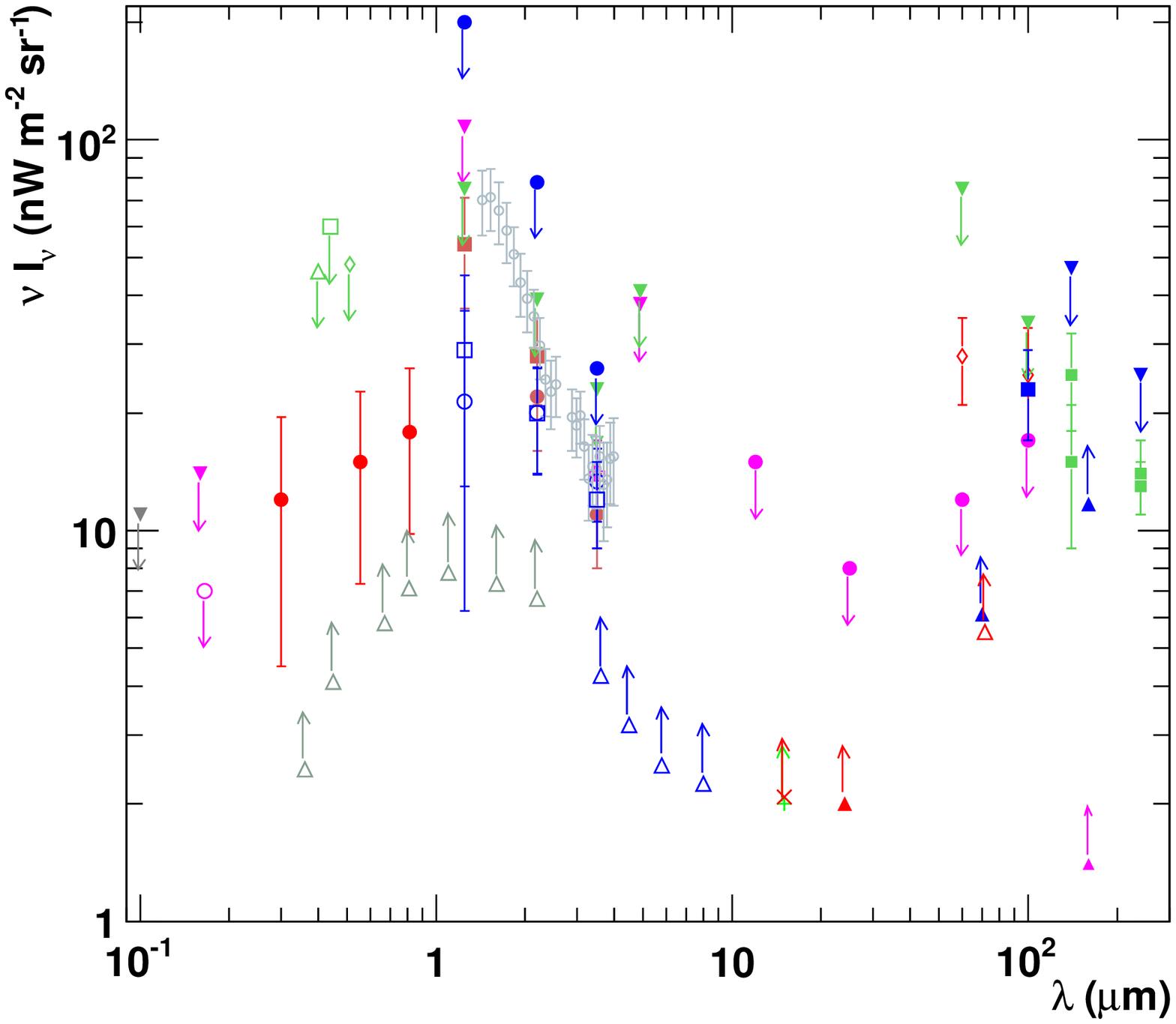}
\end{minipage}%
\begin{minipage}[c]{0.4\textwidth}
\caption{EBL measurements and limits.
Upper limits in the UV to optical:
\citet{edelstein:2000a} (grey filled triangle),
\citet{martin:1991a} (open pink circle),
\citet{brown:2000a} (filled pink triangle),
\citet{mattila:1990a} (open green triangle),
\citet{toller:1983a} / \cite{leinert:1998a} (open green square),
\citet{dube:1979a} / \cite{leinert:1998a} (open green diamond);
Tentative detection in the UV/optical:
\citet{bernstein:2002a,bernstein:2005a} (filled red circle);
Lower limits from source counts: 
\citet{madau:2000a} (open grey triangles),
\citet{fazio:2004a} (open blue triangles),
\citet{elbaz:2002a} (green cross),
\citet{metcalfe:2003a} (red x),
\citet{papovich:2004a} (filled red triangle),
\citet{dole:2006a} (filled pink triangles),
\citet{frayer:2006a} (open red triangle);
Detections in the near IR:
\citet{dwek:1998b} (open pink cross),
\citet{gorjian:2000a} (filled brown circle),
\citet{wright:2000a} (open blue squares),
\citet{cambresy:2001a} (filled brown squares),
\citet{matsumoto:2005a} (small open grey circles),
\citet{levenson:2007a} (open blue circles);
Upper limits from direct measurements:
\citet{hauser:1998a} (filled green triangles),
\citet{dwek:1998b} (filled pink triangles),
\citet{lagache:2000a} (filled blue triangles);
Upper limits from fluctuation analysis:
\citet{kashlinsky:1996a} (filled blue circles),
\citet{kashlinsky:2000a} (filled pink circles);
Lower limits from stacking analysis in the far-IR:
\citet{dole:2006a} (blue triangles);
Detections in the far-IR:
\citet{hauser:1998a} (filled green squares),
\citet{lagache:2000a} (tentative, filled blue square),
\citet{finkbeiner:2000a} (tentative, open red diamonds).
}
\end{minipage}
\label{Fig:EBLMeasurements01}
\end{figure*}
             
 The diffuse extragalactic background light (EBL) in the UV to far-IR
wavelength regime carries important information about the galaxy and star
formation history of the universe. The present EBL consists of the
integrated electromagnetic radiation from all epochs, which is redshifted,
corresponding to its formation epoch.  In the energy density distribution, a
two-peak structure is commonly expected: a first peak around 1$\,\mu$m produced
by starlight and a second peak at $\sim$100$\,\mu$m resulting from starlight
that has been absorbed and reemitted by dust in galaxies. Other contributions, like
emission from AGN and quasars, are expected to produce no more than 5 to 20\% of
the total EBL density in the mid IR (see e.g. \citealt{matute:2006a} and
references therein).

Solid lower limits for the EBL level have been derived from source counts
\citep[e.g.][]{madau:2000a,fazio:2004a,frayer:2006a,dole:2006a}.
Direct measurements of the EBL have proven to
be a difficult task due to dominant foregrounds mainly from inter-planetary
dust (zodiacal light) \citep[e.g.][]{hauser:1998a}, and it is not expected that
the sensitivity of measurements will greatly improve over the next few years.
Several upper limits were reported from direct observations
\citep[e.g.][]{hauser:1998a} and from fluctuation analyses
\citep[e.g.][]{kashlinsky:2000a}.  In total, the collective limits on the EBL
between the UV and far-IR confirm the expected two-peak structure, although the
absolute level of the EBL density remains uncertain by a factor of 2 to 10 (see
Fig.~\ref{Fig:EBLMeasurements01}). In addition, several direct detections have
also been reported, which do not contradict the limits but lie significantly above
the lower limits (see \citealt{hauser:2001a} and \citealt{Kashlinsky2005:EBLReview}
for recent reviews).  
In particular, a diffuse residual excess in the near IR (1 to 4\,$\mu$m) 
was reported by the IRTS satellite \citep{matsumoto:2005a},
which is significantly higher than the EBL density expected from galaxy number counts.

The reported excess lead to a controversial discussion about its origin.
If the excess were extragalactic origin (i.e. associated with the EBL),
it might be attributed to emissions by the first stars in the history of
the universe (Population III) and would make the EBL and its structure a unique
probe of the epoch of Population III formation and evolution
\citep{mapelli:2004a,kashlinsky:2005a}.  
Such a luminous Population III star generation, however, over-predicts the number of Ly-$\alpha$ emitters in ultra-deep field searches \citep{salvaterra:2006a} 
and violates common assumptions on baryon consumption and star formation rates
\citep{dwek:2005c}.
In addition, \citet{dwek:2005c} and \citet{matsumoto:2005a} point out that the 
NIR excess could be attributed to zodiacal light.

 A different approach of deriving constraints on the EBL (labeled ``clever'' by
\citealt{dwek:1994a}) became available with the detection of very high-energy
(VHE) $\gamma$-rays from distant sources \citep{punch:1992a}. These VHE
$\gamma$-rays are attenuated via pair production with low-energy photons from
the EBL \citep{gould:1967a}. It was soon realized that the measured spectra can
be used to test the transparency of the universe to VHE $\gamma$-rays and thus
to derive constraints on the EBL density \citep{fazio:1970a}. 
With reasonable assumptions about the
intrinsic spectrum emitted at the source, limits on the EBL density
can be inferred.In a pioneering work conducted
by \citet{stecker:1996a}, first limits on the EBL were derived.
The method is, however, not straightforward. It is, in
principle, not possible to distinguish between source-inherent effects (such as
absorption in the source, highest particle energies, magnetic field strength,
etc.) and an imprint of the EBL on a measured VHE spectrum.  The emission
processes in the detected extragalactic VHE $\gamma$-ray emitters\footnote{So
far, all but one detected extragalactic VHE $\gamma$-ray emitters belong to the
class of TeV blazars.} are far from being fully understood, which makes it more
difficult to use robust assumptions on the shape of the intrinsic spectrum.
Different EBL shapes can lead to the same attenuation imprint in the measured
spectra, thus a reconstructed attenuation imprint cannot uniquely be identified
with one specific EBL shape. A further uncertainty arises due to the unknown
evolution of the EBL with time, which becomes more important for distant
sources.

Nevertheless, measured VHE $\gamma$-ray spectra can provide robust upper limits
on the density and spectral distribution of the EBL, if conservative
assumptions about the emission mechanisms of $\gamma$-rays are considered. A
review of the various efforts to detect the EBL or to derive upper limits via
the observation of extragalactic VHE $\gamma$-ray sources up to the year 2001
can be found in \citet{hauser:2001a}. The measured spectrum of the TeV blazar H\,1426+428 at a redshift of $z = 0.129$ \citep{aharonian:2002a} led to a first tentative detection of an imprint of the EBL in a VHE spectrum \citep{aharonian:2002a,aharonian:2003a}. Using the H\,1426+428 spectrum, together with the spectra of previously detected TeV blazars, limits on the EBL were derived \citep{costamante:2003a,kneiske:2004a}. Later, \citet{dwek:2005a} utilized a
large sample of TeV blazar spectra and solid statistical methods to test a set
of EBL shapes on their physical feasibility. The EBL shapes were considered
forbidden, when, under the most conservative assumptions, the intrinsic spectra
showed a significant exponential rise at high energies. Using the VHE spectra of the TeV blazars Mkn\,421, Mkn\,501, H\,1426+428, and PKS\,2155-304, \citet{dwek:2005b} argued that the claimed NIR excess is very unlikely to be of extragalactic origin and that an EBL density on the level of the source counts gives a good representation of the intrinsic spectra. Recently, the detection of the two distant TeV blazars H\,2356-309 and 1ES\,1101-232 by the H.E.S.S. experiment has been used to derive strong limits on the EBL density around 2$\,\mu$m
\citep{aharonian:2006:hess:ebl:nature}. The method is based on the hypothesis
that the intrinsic TeV blazar spectrum cannot be harder than a theoretical
limit and that the spectrum of the EBL follows a certain shape
\citep{primack:1999a}. To derive limits on the EBL density, the shape is scaled
until the de-absorbed TeV blazar spectrum meets the exclusion criteria.%
\footnote{A similar technique (scaling of a certain shape until the de-abs~rbed
spectrum reaches an exclusion criterion) was previously used by
\citet{guy:2000a} to derive limits on the EBL from the Mkn~501 spectrum
measured by the CAT experiment during a flare in 1997.} The derived upper
limits are only a factor of two above the lower limits from the integrated
light of resolved galaxies.

In the past 2-3 years, many new TeV blazars have been discovered and the
established ones were remeasured with higher sensitivity with the new
Imaging Atmospheric Cherenkov Telescopes (IACT) such as H.E.S.S and
MAGIC. It is therefore of high interest whether the new measurements agree
with the previous limits on the EBL. Furthermore, to derive limits on the EBL 
from the data, it is important to scrutinize all available data together to obtain 
a consistent picture and to maximize the constraints. In this paper we use,
therefore, spectra from all TeV
blazars to derive upper limits on the EBL density in a wide wavelength
range from the optical to far-infrared. Moreover, a common criticism of
the EBL limits previously derived with this method is that the limits are obtained by
assuming a certain EBL model and e.g.\ scaling it, or by exploring just a few
model parameters. Since EBL models are complex and different models do not agree in
details, the derived limits become very model-dependent. In order to
avoid this dependency, we developed a novel technique of describing the
EBL number density by spline functions, which allows us to test
a large number of hypothetical EBL shapes. Our aims are to
\begin{enumerate}
 \item Provide limits on the EBL density, which do not rely on a predefined
       shape or model, but rather allow for any shape compatible with the current
       limits from direct measurements and model predications. 
 \item Treat all TeV blazar spectra in a consistent way, using simple and generic 
       assumptions about the intrinsic VHE $\gamma$-ray spectra and a statistical approach 
       to find exclusion criteria.
 \item Use spectral data from all detected TeV blazars to
       (a) derive upper limits on the EBL density from the individual spectra and then
       (b) combine these results into a single robust limit on the EBL density for a wide wavelength range.  
\end{enumerate}
 
 The paper is organized as follows: In Section~\ref{sec:grid} we describe our
method of using splines and a grid in EBL density vs. wavelength to construct different shapes, in
Section~\ref{sec:tevspectra} we introduce the TeV blazar spectra, and in
Section~\ref{sec:criteria} we describe the criteria we impose on these
spectra to derive limits on the density of the EBL. In
Section~\ref{Section:IndividualResults} results for individual sources and in
Section~\ref{Section:CombinedResults} the combined results are discussed. Last
but not least, we give a conclusion and outlook in
Section~\ref{Section:Conclusion}.
 
 Throughout the paper we adopt a Hubble constant of $H_0 =
72\,$km\,$s^{-1}$Mpc$^{-1}$ and a flat universe cosmology with a matter density
normalized to the critical density of $\Omega_{m} = 0.3$ and $\Omega_{\lambda}
= 0.7$. 

\section{Grid scan of the EBL with splines}\label{sec:grid}

\begin{figure}[t,b]
\centering
\includegraphics[width=0.45\textwidth]{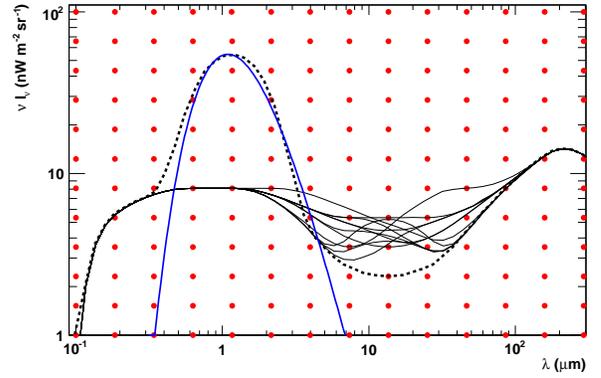}
\caption{Examples for spline shapes resulting from the grid layout, overlaid on the grid points (red filled circles). The dashed black line illustrates the thinnest structure that can be achieved with our grid setup. A Planck spectrum (blue line) is given for comparison. Further examples for shapes are given as solid black lines.}
\label{Fig:GridResolution}
\end{figure}

\begin{figure}[t,b]
\centering
\includegraphics[width=0.45\textwidth]{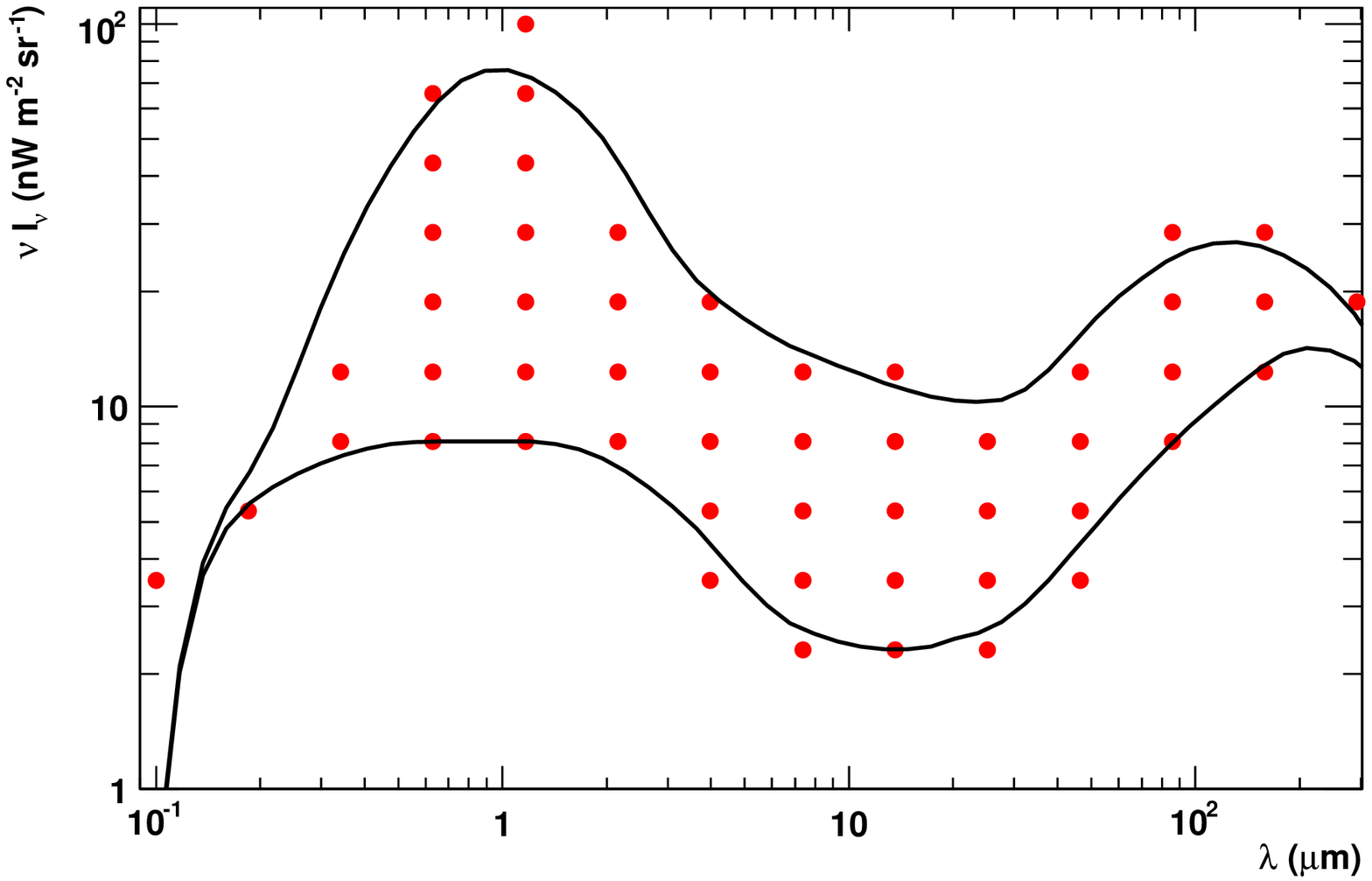}\\
\includegraphics[width=0.45\textwidth]{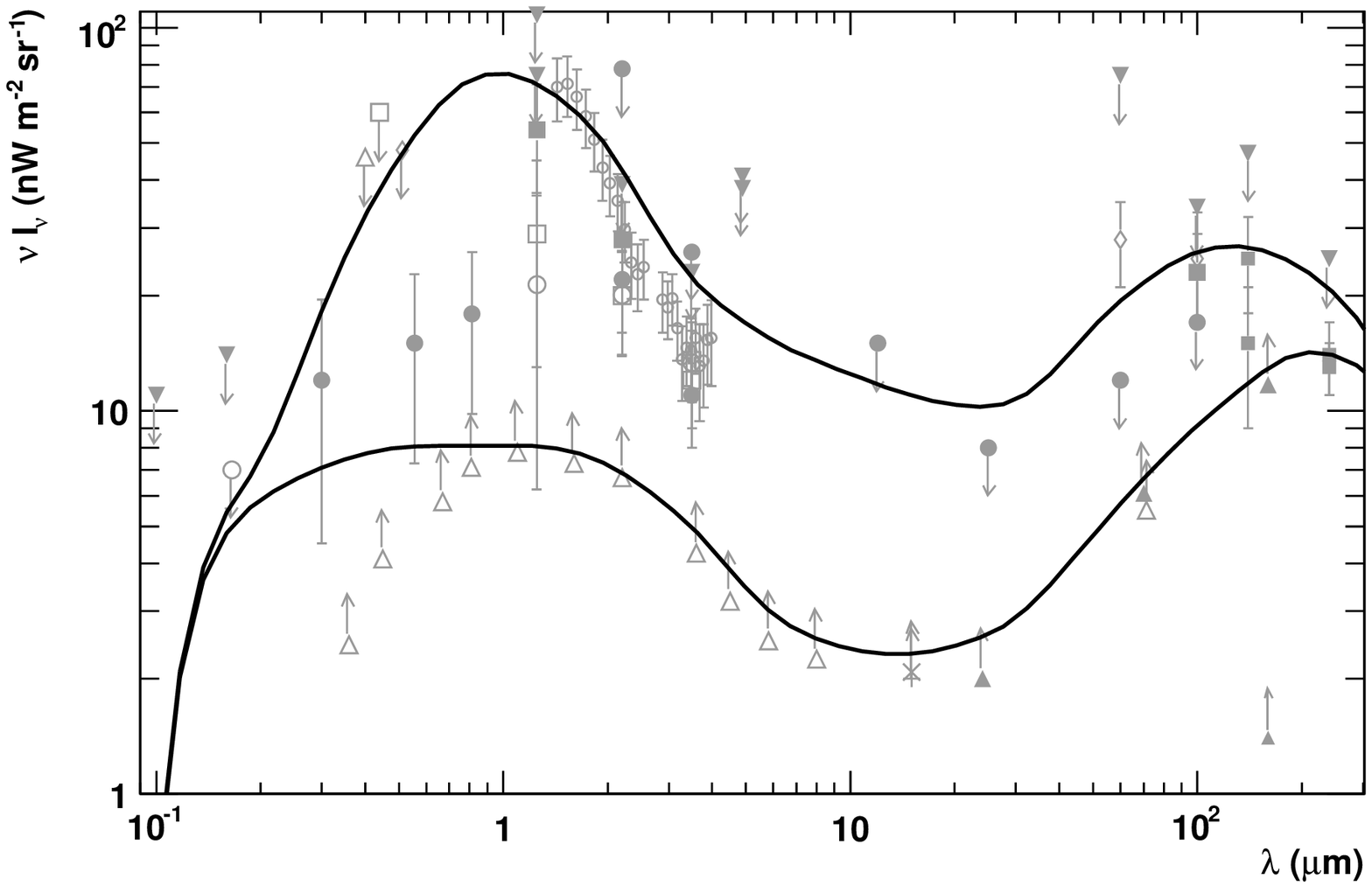}
\caption{\textsl{Top:} Grid points (red filled circles) and minimum and maximum shape of the scan. \textsl{Bottom:} Minimum and maximum shape overlaid on the EBL measurements from Fig. \ref{Fig:EBLMeasurements01}}
\label{Fig:Grid01}
\end{figure}
 
 TeV $\gamma$-rays traversing the extragalactic radiation fields are absorbed
via pair production with the low-energy photons of the EBL: $\gamma_{\rm{TeV}}
\; \gamma_{\rm{EBL}} \rightarrow e^{+} \; e^{-}$.  Details on the cross section
\citep{heitler:1960a} and the exact calculations can be found elsewhere (see
\citealt{dwek:2005a} for an overview). Here, we focus on the new technique using
splines (as introduced below) to be able to examine as many as several million
possible EBL shapes.
 
 To calculate the optical depth $\tau_\gamma$ for a TeV-$\gamma$ ray of energy
$E_\gamma$ emitted at a redshift of $z$ for one realization/shape of the EBL,
one needs to solve a three-fold integral over the distance $\ell$, the
interaction angles between the two photons $\mu = \cos \theta$ and the number
density of the EBL photons $n_\epsilon$:
\begin{eqnarray}\label{Eq:OpticalDepth} &&\tau_{\gamma}(E_{\gamma},\ z) = \\
\nonumber &&\int_0^z {\rm d}\ell ({\rm z}) \int_{-1}^{+1}{\rm d}\mu\ {1-\mu
\over 2}\int_{\epsilon'_{th}}^{\infty}{\rm d}\epsilon'\
n_{\epsilon}(\epsilon', z)\ \sigma_{\gamma \gamma}(E'_{\gamma},\ \epsilon',\ \mu)
\end{eqnarray}
where $\sigma_{\gamma\gamma}$ is the pair production cross section, $\epsilon$
the energy of the EBL photon, and $E_\gamma'$ and $\epsilon'$ refer to
redshifted energy values. For the large number of shapes ($\sim$8$\,$million)
we will analyze, a full numerical integration would require an extensive amount
of computing power. To avoid this problem, we parametrize the EBL number density at $z = 0$ as a spline
\begin{equation}\label{Eq:BSpline01} \ n_{\epsilon}(\epsilon) = \sum_{i = 0}^k
w_i s_{i, p}(\epsilon) \end{equation}
with
\begin{eqnarray}\label{Eq:BSpline02} s_{i, 0}(\epsilon) & = & \left\{
\begin{array}{l}1 \;\;\; \mbox{if} \; \epsilon_i \leq \epsilon <  \epsilon_{i +
1} \; \mbox{and} \; \epsilon_i < \epsilon_{i + 1}  \;\;\; \mbox{otherwise}\\
0\\ \end{array}\right.  \\ \nonumber s_{i, p}(\epsilon) & = &
\frac{\epsilon - \epsilon_i}{\epsilon_{i + p} - \epsilon_i} s_{i, p -
1}(\epsilon) + \frac{\epsilon_{i + p + 1} - \epsilon}{\epsilon_{i + p + 1} -
\epsilon_{i + 1}} s_{i +1, p - 1}(\epsilon)\;.  \end{eqnarray}
Here $p$ is the order of the spline and is set to $p = 3$ throughout the paper, $k$
is the number of supporting points, $\epsilon_1, ..., \epsilon_k$ are the
positions of the supporting points of the curve, and $w_1, ..., w_k$ are weights
controlling the shape of the curve.

A way to visualize this spline is to add a number of Gaussian-like base
functions to lead to a smooth curve, whereby the base functions can be
weighted ($w_i$) to achieve different overall shapes.  These base functions are
characterized by a position of the center and a certain width,
which depends on the order $p$ and the spacing of the supporting points. 

By inserting Eq.~(\ref{Eq:BSpline01}) into Eq.~(\ref{Eq:OpticalDepth})
and then swapping the integration and the summation, one obtains an expression
for the optical depth, where the integral can be solved easily for a certain redshift and set of supporting
points of the spline. The optical depth
can then be calculated by a simple summation, where the shape of the EBL is
determined by the choice of a set of weights. 

To estimate the numerical uncertainties in the integration of the base functions,
the result for $\tau_\gamma$ from the summation is compared to the results from a
full integration over the actual EBL shapes for several shapes and spectra used in this paper. 
The deviation for almost all settings is found to be less then 0.5\%.
Even for extreme cases (high redshift, high EBL density), the deviation is
always less than 2\%. These small deviations arise from 
inaccuracies in the numerical integration of the base functions.

The spline parameterization is used to construct a set of EBL shapes using a
grid in EBL energy density vs.  wavelength. The x-positions of the grid points
(wavelength) are used as positions for the supporting points $\epsilon_i$ of
the spline. The y-positions of the grid points (energy density) are used as
weights $w_i$. 

For the supporting points (x-axis of the grid) $\epsilon_i$, we use 16
equidistant points in $\log_{10}(\lambda)$ ($\lambda = h c / \epsilon$)
from 0.1 to 1000$\,\mu$m. The number and distance of the supporting points,
together with the order of the spline, determines the width of the
structures, which can be described with the spline. Given the grid setup
and the spline of order $p = 3$, the thinnest peak or dip that can be
modeled is about three grid points wide (Fig.~\ref{Fig:GridResolution}).
This minimum width is similar to a Planck spectrum
from a black body radiation (Fig.~\ref{Fig:GridResolution}).  It is expected that the EBL
originates mainly from an overlap of the redshifted spectra of
single stars, for which a Planck spectrum is a good first-order
approximation, and dust reemission. Thus, an EBL structure thinner than a Planck spectrum is
unlikely.

Noteworthy, any extremely sharp and strong cut-offs or bumps can
only be described in an approximate way. Such features can arise from the
Lyman-$\alpha$ drop-off of massive Population~III stars in certain models
\citep[e.g.][]{salvaterra:2003a}, but they are generally not expected on
larger wavelength scales, since they would be smoothed out by redshift. Of
course the upper limits depend on the thickness of the structures modeled,
and the choice of minimum thickness has to be physically motivated, as it
is in our case.

The weights $w_i$  of the spline (y-axis of the grid) are varied in 12
equidistant steps in $\log_{10}$(energy~density) from 0.1 to
100$\,$nW$\,$m$^{-2}\,$sr$^{-1}$. Since we apply the grid point positions
directly as weights $w_i$, the resulting curve does not directly go through the
grid points. So for all limits, etc., the actual curve has to be considered and not the
grid point positions. Our use of the grid point positions as weights also results in
several shapes lying between two grid points. This is illustrated in
Fig.~\ref{Fig:GridResolution} for two arbitrarily chosen grid points between 10
and 20$\,\mu$m.

For the scan, a subset of grid points is selected such that all resulting shapes
are within the limits given by the galaxy counts on the lower end and by the
limits from the direct measurements and the fluctuation analyses on the upper
end.  In the 20 to 100$\,\mu$m regime, we choose a somewhat looser shape, given
the wider spread of the (tentative) measurements (Fig.~\ref{Fig:Grid01}).
By iterating over all grid points,
we obtain 8$\,$064$\,$000 different EBL shapes, which will be examined.

The generic shapes used in this analysis have no redshift dependency, thus
evolution of the EBL cannot be taken into account. We assume a constant photon number density only expanding and shifting in wavelength with expansion of the universe.
 Neglecting the evolution of the EBL is a valid assumption for nearby sources (e.g. Mkn 501), but for
distant sources this can result in an additional error in the order of 10 to
20\% depending on the evolution model and redshift (see e.g.
\citealt{aharonian:2006:hess:ebl:nature}). The possible systematic error arising
from this assumption will be discussed further in Sect.~\ref{Section:Conclusion}.

For a given optical depth $\tau_{\gamma}(E_{\gamma},\,z)$ one can determine the 
intrinsic differential energy spectrum using
\begin{equation}
   dN/dE_{\mathrm{intr.}}\,=\, dN/dE_{obs} \times \exp[\tau_{\gamma}(E_{\gamma},\,z)],
\end{equation}
where $dN/dE_{obs}$ is the observed spectrum. The intrinsic spectrum is then
tested on its physical feasibility as described in Section~\ref{sec:criteria}.
This process is repeated for all VHE $\gamma$-ray sources (described in the
next section) and for all 8$\,$064$\,$000 EBL shapes.

\section{TeV blazar sample}\label{sec:tevspectra}

\begin{table*}[t] \begin{center} \label{tab:AGNspec} 
\caption{TeV blazar spectra used in this paper.\label{Table:tevspectra}}
\begin{tabular}{ l l l c l c l }
\hline \hline
Source & Redshift & Experiment & Energy range  & \mc{Slope}                      & Cut-off energy       & Reference \\
       &          &            &  (TeV)    & \mc{$\Gamma\pm\sigma_{\mathrm{st}}\pm\sigma_{\mathrm{sy}}$} & (TeV) &  \\
\hline 
\object{Mkn\,421}   & 0.030   &  MAGIC      &  0.10 -- 3.0  & $2.20 \pm0.08\pm0.20 $ & $1.44\pm0.28$ & \citet{albert:2006:magic:mkn421} \\
Mkn\,421   & 0.030   &  HEGRA      &  0.70 -- 18.0 & $2.19 \pm0.02\pm0.20 $ & $3.6+0.4-0.3$ & \citet{aharonian:2002e} \\
Mkn\,421   & 0.030   &  Whipple    &  0.35 -- 0.90 & $2.31 \pm0.04\pm0.05 $ & ------        & \citet{krennrich:2002a} \\
\object{Mkn\,501}   & 0.034   &  HEGRA      &  0.50 -- 22.0 & $1.92 \pm0.03\pm0.20 $ & $6.2 \pm 0.4$ & \citet{aharonian:1999b} \\
\object{1ES\,2344+514} & 0.044 &  Whipple  &  0.80 -- 11.0 &  $2.54 \pm0.17\pm0.07 $ & ------  & \citet{schroedter:2005a} \\
\object{Mkn\,180}   & 0.045   &  MAGIC      &  0.14 -- 1.5  & $3.25 \pm0.66\pm0.20 $ & ------  & \citet{albert:2006:magic:mkn180} \\
\object{1ES\,1959+650} & 0.047   &  HEGRA  &  1.5 -- 13.0  & $2.83 \pm0.14\pm0.08 $ & ------        & \citet{aharonian:2003c} \\
\object{PKS\,2005-489} & 0.071   & H.E.S.S.&  0.20 -- 2.5  & $4.0 \pm0.4\,(\pm0.2)$ & ------        & \citet{Aharonian2005:HESS.PKS2005} \\
\object{PKS\,2155-304} & 0.116   & H.E.S.S.&  0.20 -- 3.5   & $3.37\pm0.07\pm0.10 $ & ------        & \citet{Aharonian2005:HESS.PKS2155.MWL} \\
\object{H\,1426+428}   & 0.129   & HEGRA   &  0.70 -- 12.0  & $2.6 \pm0.6\pm0.1   $ & ------        & \citet{aharonian:2003a} \\
\object{H\,2356-309}   & 0.165   & H.E.S.S.&  0.16 -- 1.0   & $3.06\pm0.21\pm0.10 $ & ------        & \citet{Aharonian2006:HESS.H2356} \\
\object{1ES\,1218+304} & 0.182   & MAGIC   &  0.08 -- 0.7   & $3.0 \pm0.4 \pm0.6  $ & ------        & \citet{albert:2006:magic:1ES1218} \\
\object{1ES\,1101-232} & 0.186   & H.E.S.S.&  0.16 -- 3.3   & $2.88\pm0.14\pm0.1  $ & ------        & \citet{aharonian:2006:hess:ebl:nature} \\
\hline
\end{tabular}
\end{center}
\end{table*}

\begin{figure*}[t]
\centering
\includegraphics[width=\textwidth]{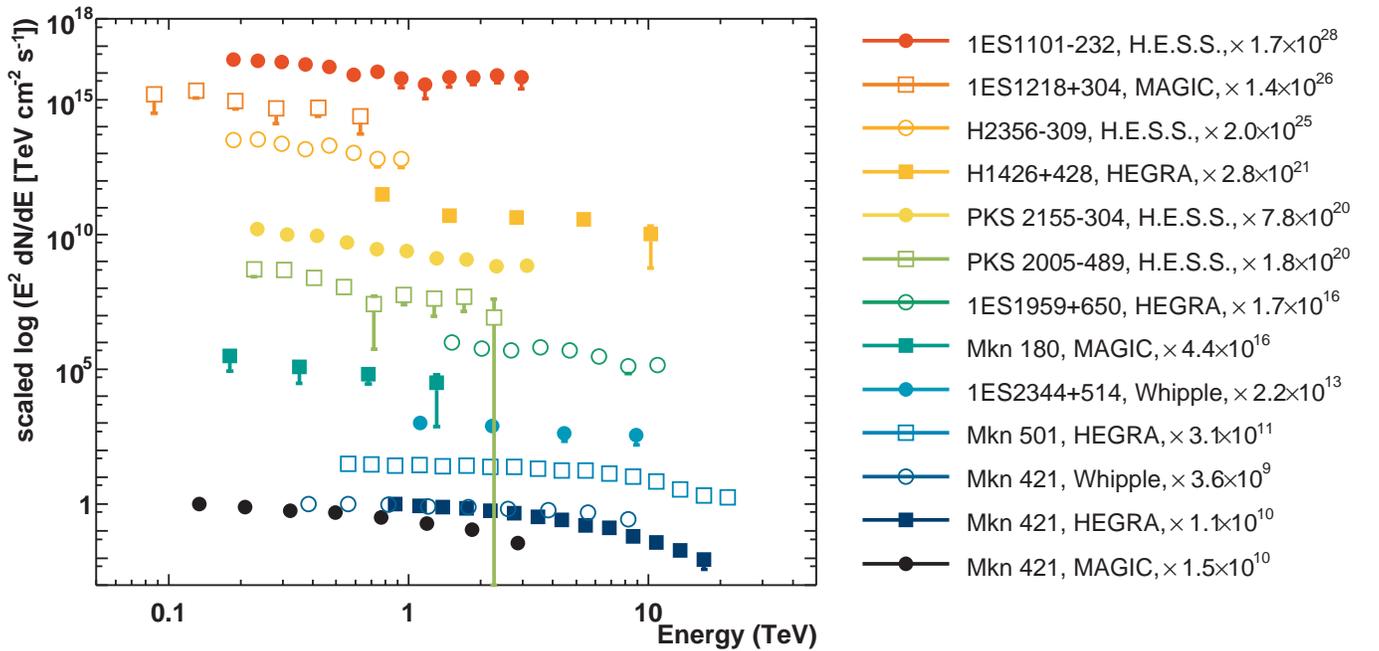}
\caption{TeV blazar sample chosen for this study. From all detected TeV blazars, at least one spectrum is shown with the exception of PG\,1553+113 (unknown redshift). The spectra are multiplied by E$^2$ to emphasize spectral differences and are spread out along the Y-axis and ordered in redshift to avoid cluttering of the plot (the corresponding scaling factors are given in the legend of the plot).}
\label{Fig:AGNspectra}
\end{figure*}

Since the detection of the first extragalactic VHE $\gamma$-ray source in 1992
\citep{punch:1992a}, a wealth of new data from this source type has become
available. Up to now all detected VHE $\gamma$-ray sources belong to the class
of active galactic nuclei (AGN) and, with the only one exception of the radio
galaxy M\,87 \citep{aharonian:2003b,aharonian:2006:hess:m87:science}, to the
subclass of TeV blazars \citep{horan:2004a}. AGNs are known to be highly
variable sources, with variations of the absolute flux levels in the TeV range by more
than an order of magnitude and on time scales as short as 15\,min
\citep[e.g.][]{gaidos:1996a,aharonian:2002e}. Changes in the spectral shapes
have also been observed \citep[e.g.][]{krennrich:2002a,aharonian:2002e}.

In this study we utilize spectral data obtained during the past seven years by
four different experimental groups operating ground-based imaging atmospheric
Cherenkov telescopes (IACTs): Whipple \citep{finley:2001a}, HEGRA
\citep{daum:1997a}, H.E.S.S. \citep{Hinton2004:HESSStatus} and MAGIC
\citep{cortina:2005a}. We select at least one spectrum for every extragalactic
source with known redshift. If there is more than one measurement with a
comparable energy range, we take the spectrum with the better statistics and the
harder spectrum (expected to give stronger constraints, see
Sect.~\ref{sec:criteria}). If different measurements of one source cover
different energy ranges, we include both spectra as independent tests. Another
possibility would be to combine the measurements from different experiments, as
was done before in the case of  H\,1426+428 \citep{aharonian:2003a}.
However, since the sources are known to be variable in the flux level and
spectral shape, a combination of non-simultaneous data is not trivial. In
addition, in a conservative approach one has to consider the
systematic errors reported by the individual experiments, which are around 20\%. Our method is quite sensitive to the errors of the flux, and this
additional error would weaken our results. We therefore do not use combined
spectra. The selected TeV blazar spectra are summarized in
Fig.~\ref{Fig:AGNspectra} and Table~\ref{Table:tevspectra}.
In the case the measured spectrum
can be described well by a simple power law $dN/dE \propto E^{-\Gamma}$, only
the slope $\Gamma$ is quoted. Otherwise, the cut-off energy $E_{\mathrm{coff}}$
according to $dN/dE \propto E^{-\Gamma}\,\exp(-E/E_{\mathrm{coff}})$ is also
quoted. If no systematic error in the photon index is given in the corresponding
paper, we use a value of 0.2 (values in brackets).
In Fig.~\ref{Fig:AGNspectra} the spectra are multiplied by E$^2$ to emphasize
spectral differences and are spread out along the Y-axis and ordered in
redshift to avoid cluttering the plot.

We did not include data from the radio galaxy M\,87, which was recently
confirmed as a VHE $\gamma$-ray emitter
\citep{aharonian:2006:hess:m87:science}. M\,87 is a nearby source ($z =
0.00436$), so even for high EBL densities the attenuation is weak and would
only be noticeable at high energies $\sim$30\,TeV. Nevertheless, M\,87 has a
hard spectrum with a photon index of $\Gamma \sim 2.2$ currently measured up to
20 TeV, so further observations that extend the energy range to even higher
energies could make M\,87 an interesting target for EBL studies as well.

The cross section $\sigma_{\gamma \gamma}$ of the pair production
process has a distinct peak close to the threshold. The maximum attenuation
of VHE photons of energy $E_{\gamma}$ occurs by interaction with low-energy
photons with a wavelength
\begin{equation}\label{Formula:EBLRange}
 \lambda^{\mathrm{max}} (\mu \mathrm{m})  \approx 1.24\; E_{\gamma} (\mathrm{TeV}) .
\end{equation}
Since the selected spectra cover an energy region from 100\,GeV up to more than
20\,TeV, the EBL wavelength range for the absorption of VHE $\gamma$-rays spans
from UV ($\sim$0.1$\,\mu$m) to the mid IR ($\sim$30$\,\mu$m). This region
is of particular cosmological interest since it might contain a signature of
Population III stars.


\section{Exclusion criteria for the EBL shapes}\label{sec:criteria}

\begin{table*}[h]
\begin{center}
\caption{Analytical functions that are
used to fit the intrinsic TeV blazar spectrum.}
\begin{tabular}{ c p{5cm} c l l }
\hline \hline
\# &   Description                     & Abbreviation   & Formula $ f(E) = dN/dE$ & Parameters to evaluate \\
\hline
1      & simple power law            &  PL      &  $N_0 E^{-\Gamma} $  & $\chi^{2},\, \Gamma^{\mathrm{PL}}$ \\ 
2 & broken power law with transition region & BPL &  
             $N_0 E^{-\Gamma_1} \left[1+\left(\frac{E}{E_b}\right)^f \right]^{\frac{\Gamma_1-\Gamma_2}{f}} $ 
                                                         & $\chi^{2},\, \Gamma_{1}^{\mathrm{BPL}}, \, \Gamma_{2}^{\mathrm{BPL}}$ \\
3      & broken power law with transition region and super-exponential pile-up &  BPLSE  &  
   $N_0 E^{-\Gamma_1} \left[1+\left(\frac{E}{E_b}\right)^f \right]^{\frac{\Gamma_1-\Gamma_2}{f}} 
             \exp\left(\frac{E}{E_{\mathrm{p}}}\right) $  
 & $\chi^{2} $ \\
4 & double broken power law with transition regions     &DBPL & 
   $N_0 E^{-\Gamma_1} \left[1+\left(\frac{E}{E_{b1}}\right)^{f_1} \right]^{\frac{\Gamma_1-\Gamma_2}{f_1}} 
     \left[1+\left(\frac{E}{E_{b2}}\right)^{f_2} \right]^{\frac{\Gamma_2-\Gamma_3}{f_2}} $   
                & $\chi^{2}, \,\Gamma_{1}^{\mathrm{DBPL}},\,\Gamma_{2}^{\mathrm{DBPL}},\,\Gamma_{3}^{\mathrm{DBPL}}$ \\  \\
5      & double broken power law with transition regions and super-exponential pile-up &  DBPLSE &  DBPL $\times \exp \left(\frac{E}{E_{\mathrm{p}}}\right) $   &  $\chi^{2} $ \\
\hline
\end{tabular} \label{tab:functions} 
\end{center}
\end{table*}

We aim to construct an upper limit on the EBL density using the TeV blazar sample
described in Sect.~\ref{sec:tevspectra}. In order to achieve this, we examine
every EBL shape (as introduced in Sect.~\ref{sec:grid}) to see whether the
intrinsic TeV blazar spectra, which result from 
correcting the measured spectra for the corresponding optical depths, are physically feasible. EBL shapes
are considered to be allowed if the intrinsic spectra of all tested TeV
blazars are feasible.  As an upper limit we define the upper envelope of all
allowed shapes.  It is constraining in the wavelength range where it lies below
the maximum shape of the scan. Otherwise no upper limit is quoted. 

There are different ways to examine a TeV blazar spectrum upon its feasibility.
In this paper we follow very general arguments arising from the shock
acceleration scenario of relativistic particles. In this well-accepted view,
electrons are Fermi-accelerated with a resulting power-law spectrum of $dN/dE
\sim E^{-\alpha}$, with a slope $\alpha$ of about 2. Due to a faster cooling of
high energy electrons compared to lower energies, the slope $\alpha$ can be
steeper than 2, but there is no simple theoretical possibility to produce an
electron spectrum with a harder spectrum. Given an electron spectrum, one can
calculate the slope of the synchrotron energy spectrum: it is $dN/dE \sim
E^{-\Gamma}$ with a photon index $\Gamma = \frac{\alpha+1}{2} = 1.5$.  The
energy spectrum of inverse-Compton photons, independent of the origin of the
target photons, has approximately the same photon index as the synchrotron
energy spectrum if the scattering occurs in the Thomson regime. In the case of the
Klein-Nishima regime, the index is even larger. Thus, the photon index of the
energy spectrum of VHE photons originating from an IC scattering is $\Gamma =
1.5$ or larger.  In the case of a hadronic origin, their spectrum is more
complicated.  Yet, the resulting VHE photons originate from 
pion decays leading to a photon spectrum with $\Gamma = 2$, which is softer
than a possible photon spectrum with a leptonic origin. In conclusion, we
assume that the photon index $\Gamma$ of the intrinsic TeV blazar spectrum is
1.5 or larger. These arguments were recently used in
\citet{aharonian:2003a,aharonian:2006:hess:ebl:nature,
aharonian2006:hess:pg1553} and \citet{albert:2006:magic:pg1553}.

However, a possibility of obtaining even harder photon spectra is not fully
excluded.  For instance, though contrary to the wide acceptance of these
general arguments, \citet{katarzynski:2006a} argue that synchrotron emission, as
well as IC scattering of relativistic electrons, does not necessarily occur
close to the region of electron acceleration.  If so, 
the electron spectrum can become truncated due to propagation
effects; i.e. the minimum energy
of electrons can be as high as several GeV. In an extreme case, we deal with a
monoenergetic spectrum of VHE relativistic electrons. Then, and this
is the most extreme case, a resulting IC photon index can be as small as
$\Gamma = 2/3$. We use this limit in the present paper, in addition to the
standard limit of $\Gamma = 1.5$, to demonstrate the strength of our method.

In the simplest models, it is assumed that VHE photons originate from a single compact region
(so-called one-zone scenario). This result in a smooth, convex spectral energy-distribution 
with two peaks at certain energies in a $\log(\nu F(\nu)) \mathrm{vs.}
\log(\nu)$ representation. Yet there are no obvious arguments against
scenarios including some kind of multizone models, which can be naturally
realized in the jets of TeV blazars. Then, the measured spectrum of VHE $\gamma$-ray
emitting sources will be  a superposition of several one-zone emission regions.
So far, there has been no indication of this in the measured spectra,
however, the attenuation of VHE $\gamma$-rays by EBL photons could hide
such substructures.

In addition to the constraints on the photon index of the intrinsic TeV blazar
spectra, we also argue that a super-exponentially rising energy spectrum with an
increasing energy is not realistic. The so-called {\it pile-up} at high energies was first
noticed by \citet{protheroe:2000a} for the Mkn 501 spectrum, and early attempts to avoid it invoked violation of the Lorentz invariance \citep{kifune:1999a,protheroe:2000a}. On the
other hand, \citet{stecker:2001a} argue that the same Mkn\,501 spectrum
data can be used to place severe constraints on the Lorentz-invariance breaking parameter.
Another possibility for explaining  pile-ups would require an ultrarelativistic jet with a very high bulk-motion Lorentz factor $\Gamma_0>3 \times 10^7$ \citep{aharonian:timokhin:2001a}. The
pile-ups, however, seem to arise at different energies for different sources,
which is not expected in these models. Moreover, the pile-ups can be easily
avoided by choosing a sufficiently low level of the EBL density. 

Based on the arguments above and including possible multizone emission scenarios,
we assume that at least one of the following smooth, analytical functions can
describe the intrinsic spectrum satisfactory well: a simple power law (PL), a
broken power law with a transition region (BPL), a broken power law with a
transition region and a super-exponential pile-up (BPLSE), a double broken
power law with two transition regions (DBPL), or a double broken power law with
two transition regions and a super-exponential pile-up (DBPLSE). The functions
are summarized in Table~\ref{tab:functions}.

In our approach of examining the  intrinsic spectra of TeV blazars, we adopt a
general assumption that one of the smooth functions above describes
the intrinsic spectrum satisfactorily. In case we fail to find
such a function, we abandon any further examination of the fit parameters, and
a given EBL shape (realization) is not excluded.  It is noteworthy that less
than 0.06\% of all intrinsic spectra of the TeV blazars could not be fitted
well by the chosen functional forms, which are described below.

In order to determine a good analytic description, we fit the intrinsic
spectrum with the functions listed above, starting with the simplest one (PL).
As ``good description'' we take a fit with a chance probability
$P_{\mathrm{Fit}} >$~5\% based on its $\chi^2$ value. The determined fit
parameters are examined to be physically meaningful as described below.  If
the fit has a probability $P_{\mathrm{Fit}} <$~5\%, we do not consider the function and examine the next function from the list (Table~\ref{tab:functions}).
Also if the fit has a chance probability $P_{\mathrm{Fit}} >$~5\%,  but
the determined parameters do not lead to an exclusion of the corresponding EBL shape, 
we take the next function from the list. The reason is that a function with more
free parameters can lead to a significantly better fit. In
order to make sure that a more complicated function indeed describes the
intrinsic shape better than a simpler one, we use the likelihood ratio test
(\citealt{eadie:1998a}, Appendix A). We require at least a 95\% probability to
prefer a function with a higher number of free parameters. Only if the function
is preferred, we examine its fitted parameters and their errors. Spectra with
$n$ data points are only fitted with functions with up to $n -1$ parameters. In
case of spectra with low statistics, we fix the softness of the break 
($f$, see Table~\ref{tab:functions}) between two spectral indices to 
an {\it a priori} chosen value of $f=4$ to allow for a fit by a
higher-order function (e.g. in case of the BPL for H\,1426+428).

We finally define the following criteria to exclude an assumed EBL shape:
\begin{itemize}
 \item At least one of the determined photon indices from the best hypothesis is outside of the allowed range, i.e.
       $\Gamma^{i,\mathrm{max}} < \Gamma^{\mathrm{limit}}$, where $\Gamma^{\mathrm{limit}} = 1.5$ or 2/3, and 
       $\Gamma^{i,\mathrm{max}} = \Gamma^i + \sigma_{\Gamma^i} + \sigma_{\mathrm{sy}}$ with
       $\Gamma^i$ and  $\sigma_{\Gamma}$ the fitted photon index and its error, respectively, and
       $\sigma_{\mathrm{sy}}$ as the systematic error on the spectral slope (as given for the corresponding
       measurement). 
 \item The best fit is obtained by one of the two shapes with an exponential pile-up (BPLSE or DBPLSE).
\end{itemize}

A single confidence level for the derived upper limits on the EBL density
cannot be quoted easily: For the test on the photon index we use a 1\,$\sigma$
confidence level (as defined for two-sided distributions, including systematic
errors). For the likelihood ratio test, we use a 2\,$\sigma$ level (as defined
for one-sided distributions). Thus, the confidence level for the upper limit
ranges from 68\% to 95\%.
 
As discussed in this section, the theoretical expectations on the smallest
possible photon index $\Gamma^{\mathrm{limit}}$ have a given spread. Thus, we
perform two EBL scans, assuming a limit of
$\Gamma^{\mathrm{limit}} =1.5 $  for the first {\it realistic} case and for the second {\it extreme} case a limit
of $\Gamma^{\mathrm{limit}} =2/3 $. From here on, we refer to {\it realistic}
and {\it extreme} scan accordingly.

Since our exclusion criteria consider the effect of an EBL shape solely on the hardness of the spectra of extragalactic VHE $\gamma$-ray sources, some shapes might be considered viable in our scan, even though other criteria for exclusion could be found (e.g. exclusion based on spectral shape of the EBL, related to the energy density of star and dust emission as discussed in \citealt{dwek:2005a}). In this respect our approach is conservative.


\section{Results for individual spectra}
\label{Section:IndividualResults}

\begin{figure}[t,b]
\centering
\includegraphics[width=0.45\textwidth]{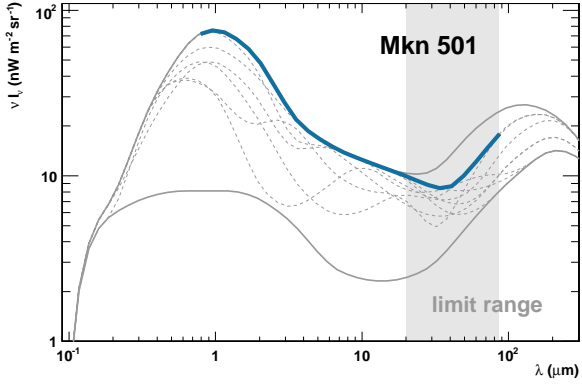}
\caption{Limits on the EBL density from Mkn\,501 ($\Gamma_{\mathrm{max}} >
1.5$). Grey solid curves are the minimum and maximum shapes of the scan; grey dashed curves are all the highest allowed EBL shapes for Mkn\,501; the thick colored curve is the corresponding envelope shape. The grey shaded area marks the wavelength region, in which the envelope shape is constraining the EBL density.} 
\label{Fig:LimitMkn501} 
\end{figure}

\begin{figure}[t,b]
\centering
\includegraphics[width=0.45\textwidth]{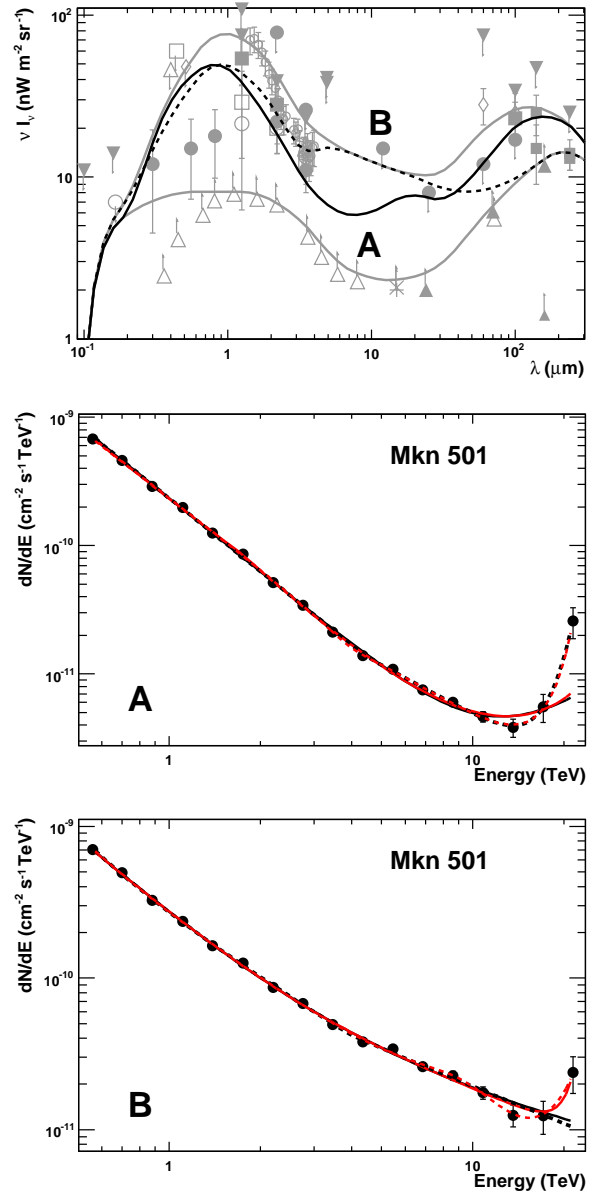}
\caption{Mkn\,501 individual results ($\Gamma_{\mathrm{max}} > 1.5$).  
\textit{Upper Panel:} Two examples for the highest allowed shapes (A - solid line; B - dashed line) overlayed on EBL measurements and the minimum and maximum shapes of the scan (grey).
\textit{Middle Panel:} Intrinsic spectrum for shape A and fit functions: BPL (solid black line), BPLSE (solid red line), DBPL (dashed black line), and DBPLSE (dashed red line). The fit parameters can be found in Table~\ref{Table:resmkn501}. The BPL and BPLSE are not good fits;
the DBPL is a good fit and its parameters are within the allowed range; the DBPLSE is not prefered over the BPL and the DBPL.
\textit{Lower Panel:} Intrinsic spectrum for shape B and the same fit functions as in the Middle Panel. The fit parameters can be found in Table~\ref{Table:resmkn501}. All four functions are good fits. No function is preferred over the BPL.} 
\label{Fig:ResultsMkn501}
\end{figure}

\begin{table}[t] 
\begin{center}
\caption{Fit results for the Mkn\,501 spectrum using the functions from
Fig.~\ref{Fig:ResultsMkn501}.}
\begin{tabular}{lccccc}
\hline \hline
& $P_{\mathrm{Fit}}$ & Pref.  & $\Gamma_{\mathrm{1,max}}$ & $\Gamma_{\mathrm{2,max}}$ & $\Gamma_{\mathrm{3,max}}$ \\
\hline
{\bf Shape A} & & & & & \\
BPL & 0.00 & --- & (2.13) & (-0.47) & ---\\
BPLSE & 0.04 & --- & --- & --- & --- \\
DBPL & 0.19 & yes & 2.12 & 1.62 & 8.92  \\
DBPLSE & 0.40 & no & --- & --- & --- \\
\hline
{\bf Shape B} & & & & & \\
BPL & 0.16 & yes  & 2.85 & 1.77 & ---\\
BPLSE & 0.23 & no & --- & --- & --- \\
DBPL & 0.12  & no & (3.91) & (1.42) & (1.05)\\
DBPLSE & 0.24 & no & --- & --- & --- \\
\hline
\end{tabular}\label{Table:resmkn501}
\end{center}
\end{table}

\begin{figure}[t,b]
\centering
\includegraphics[width=0.45\textwidth]{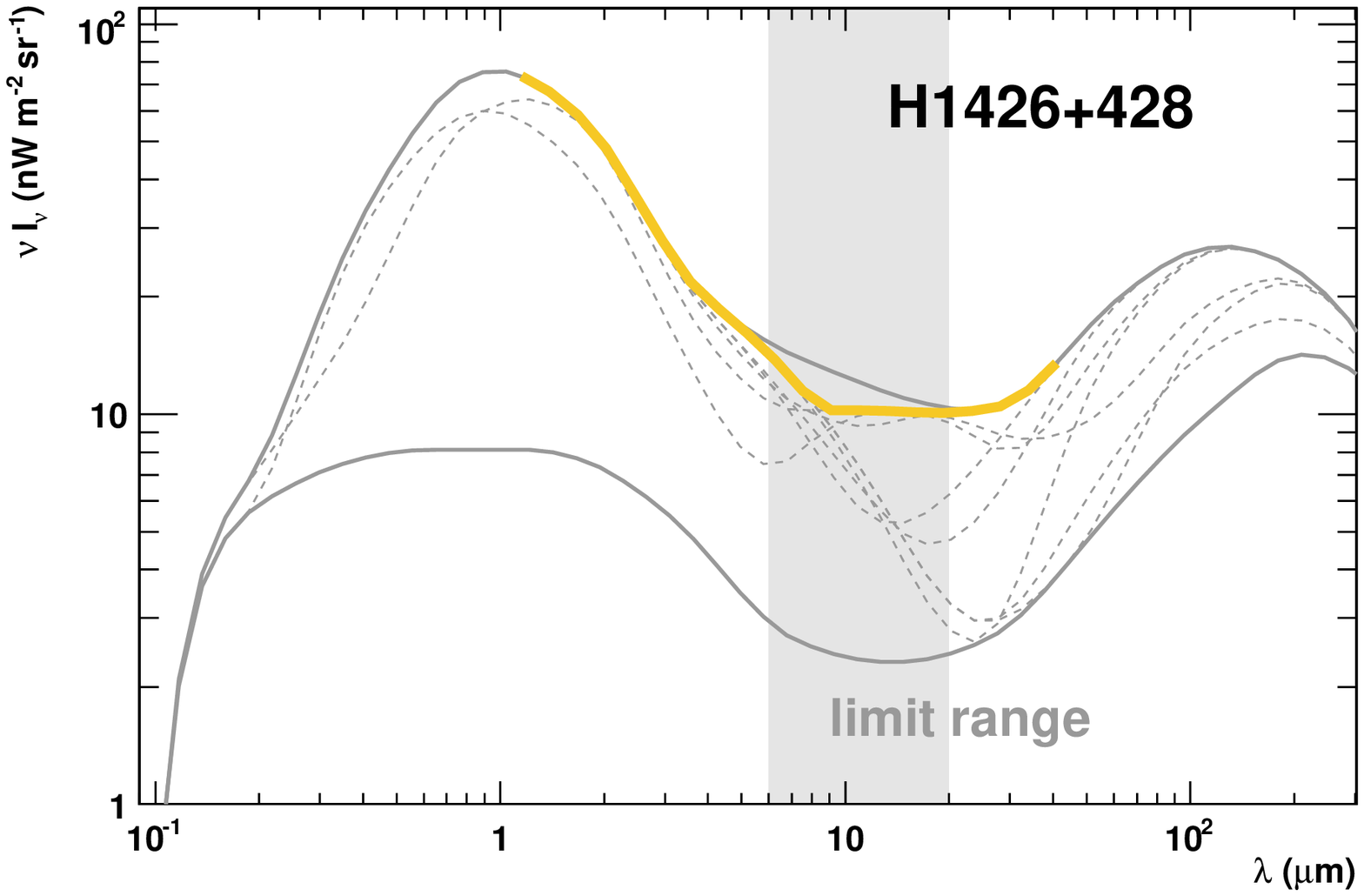}
\caption{Limits on the EBL density from H1426+428 ($\Gamma_{\mathrm{max}} > 1.5$). Grey solid curves are the minimum and maximum shapes of the scan; grey dashed curves are all the highest allowed EBL shapes for H1426+428; the thick colored curve is the corresponding envelope shape. The grey shaded area marks the wavelength region, in which the envelope shape is constraining the EBL density.}
\label{Fig:LimitH1426}
\end{figure}
 
\begin{figure}[t,b]
\centering
\includegraphics[width=0.45\textwidth]{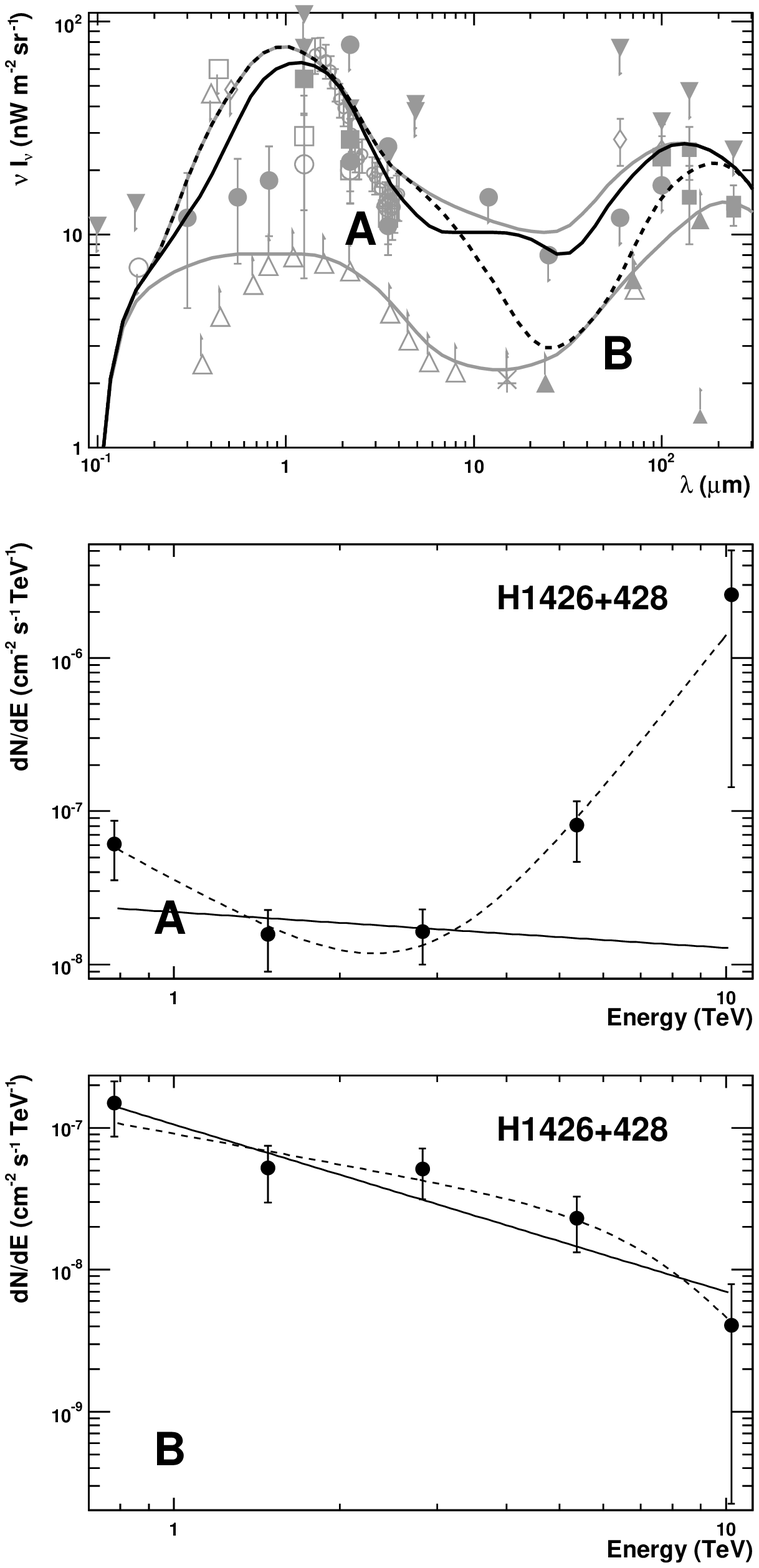}
\caption{H1426+428 individual results ($\Gamma_{\mathrm{max}} > 1.5$). \textit{Upper Panel:} Two examples for the highest allowed shapes (A - solid line; B - dashed line) overlayed on EBL measurements and the minimum and maximum shapes of the scan (grey). \textit{Middle Panel:} Intrinsic spectrum for shape A and fit functions. Both PL (solid line) and BPL (dashed line) are good fits; $\Gamma_{\mathrm{max}}^{\mathrm{PL}} =  0.23 + 1.18_{\mathrm{stat}} + 0.2_{\mathrm{sys}} = 1.61 $, and the BPL fit is preferred over the PL fit ($P_{\mathrm{(BPL\;vs.\;PL)}} = 0.97$);  $\Gamma_{\mathrm{1,max}}^{\mathrm{BPL}} =  2.01 + 1.94_{\mathrm{stat}} + 0.2_{\mathrm{sys}} = 4.15$ and $\Gamma_{\mathrm{2,max}}^{\mathrm{BPL}} =  -4.59 + 6.67_{\mathrm{stat}} + 0.2_{\mathrm{sys}} = 2.28$. \textit{Lower Panel:} Intrinsic spectrum for shape B and fit functions. Both PL (solid line)  and BPL (dashed line) are good fits; $\Gamma_{\mathrm{max}}^{\mathrm{PL}} =  1.18 + 0.24_{\mathrm{stat}} + 0.2_{\mathrm{sys}} = 1.63 $ and the BPL fit is not preferred over the PL fit ($P_{\mathrm{(BPL\;vs.\;PL)}} = 0.54$); }
\label{Fig:ResultsH1426}
\end{figure}

\begin{figure}[t,b]
\centering
\includegraphics[width=0.45\textwidth]{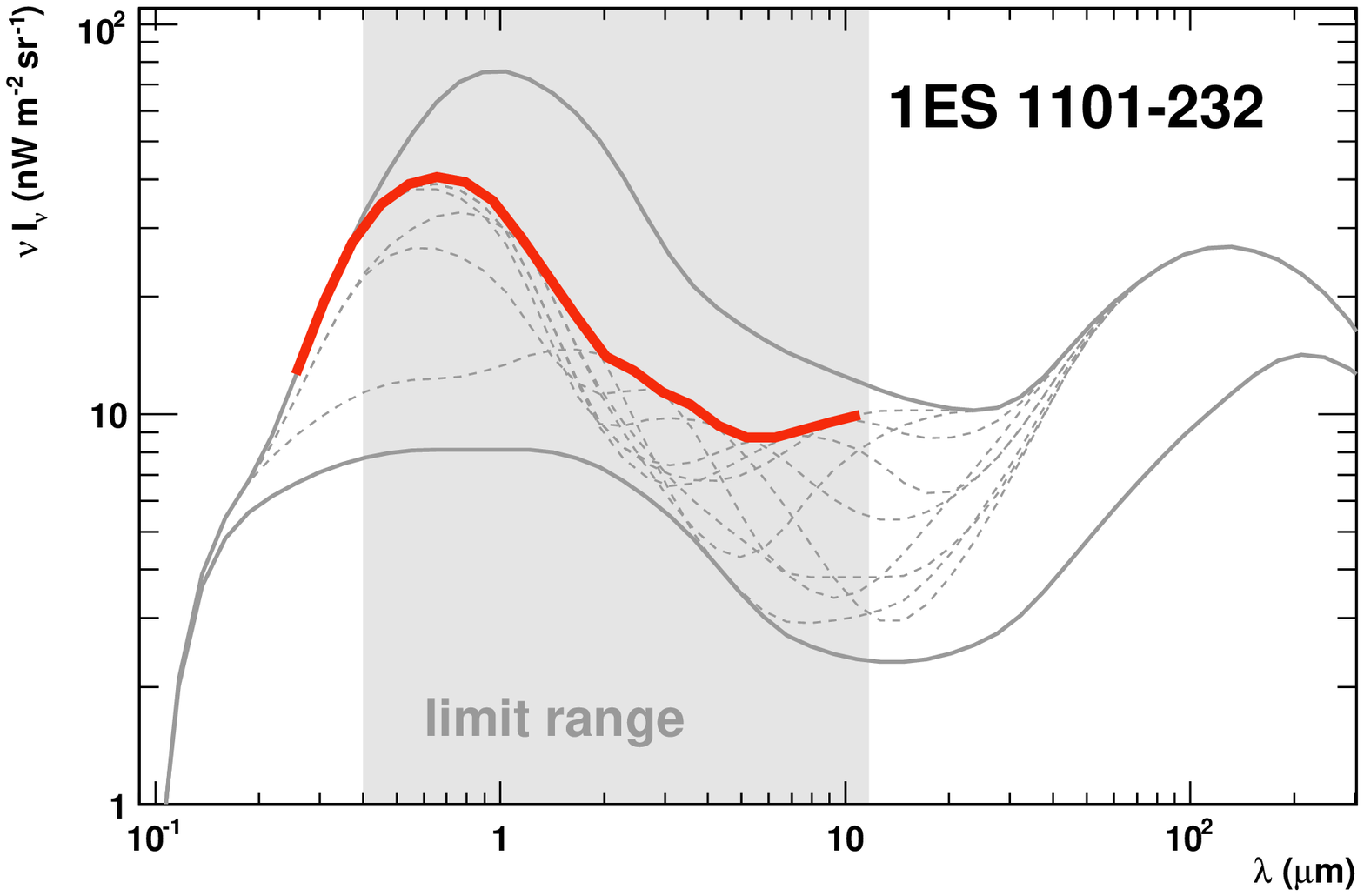}
\caption{Limits on the EBL density from 1ES\,1101-232 ($\Gamma_{\mathrm{max}} > 1.5$). Grey solid curves are the minimum and maximum shapes of the scan; grey dashed curves are all the highest allowed EBL shapes for 1ES\,1101-232; the thick colored curve is the corresponding envelope shape. The grey shaded area marks the wavelength region, in which the envelope shape is constraining the EBL density.}
\label{Fig:Limit1ES1101}
\end{figure}

\begin{figure}[t,b]
\centering
\includegraphics[width=0.45\textwidth]{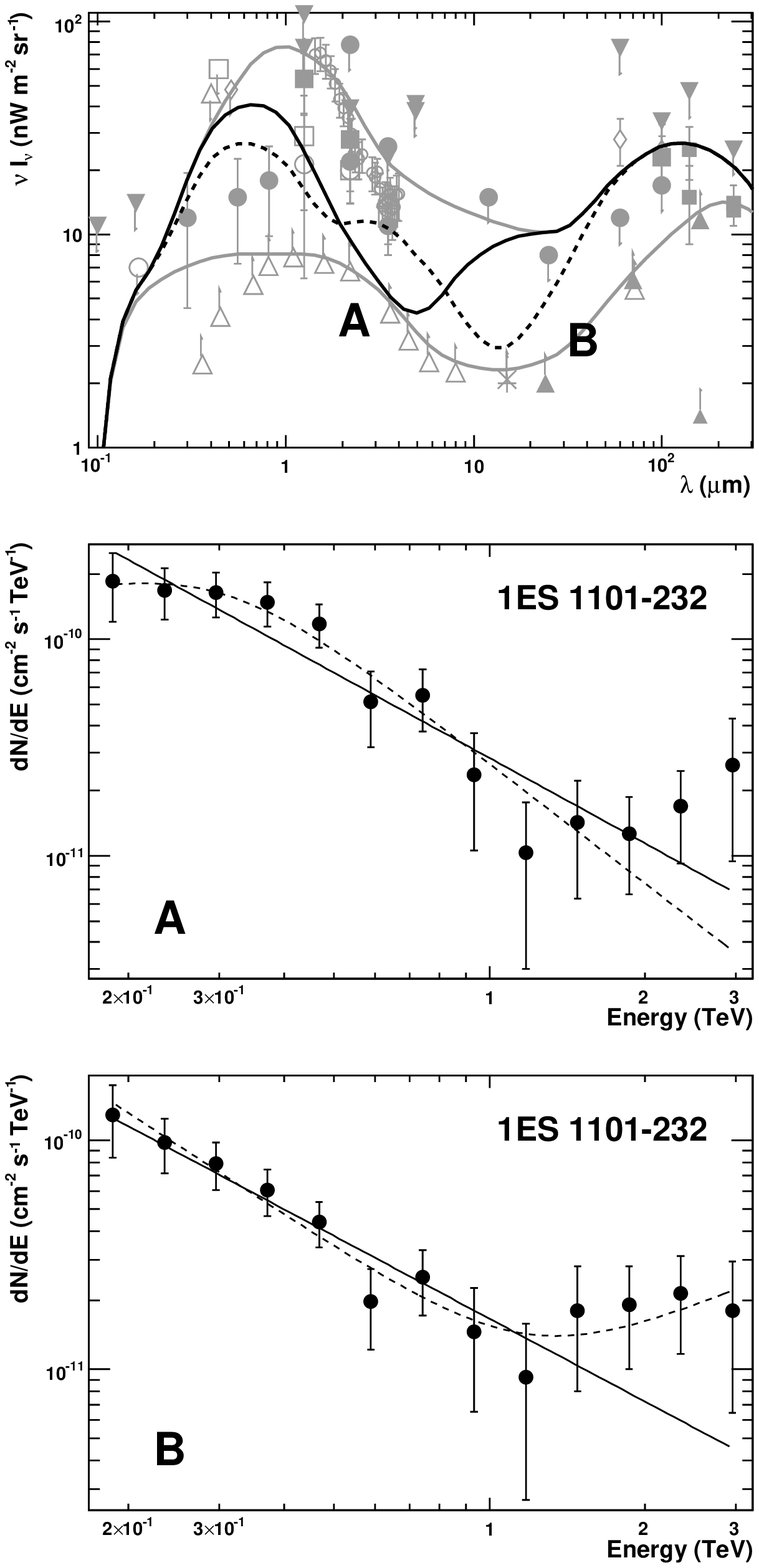}
\caption{1ES\,1101-232 individual results ($\Gamma_{\mathrm{max}} > 1.5$). \textit{Upper Panel:} Two examples for the highest allowed shapes (A - solid line; B - dashed line) overlayed on EBL measurements and the minimum and maximum shapes of the scan (grey). \textit{Middle Panel:} Intrinsic spectrum for shape A and fit functions. Both PL (solid line) and BPL (dashed line) are good fits; $\Gamma_{\mathrm{max}}^{\mathrm{PL}} = 1.31 + 0.13_{\mathrm{stat}} + 0.1_{\mathrm{sys}} = 1.53$ and the BPL fit is not preferred over the PL fit ($P_{\mathrm{(BPL\;vs.\;PL)}} = 0.87$). \textit{Lower Panel:} Intrinsic spectrum for shape B and fit functions. Both PL (solid line)  and BPL (dashed line) are good fits; $\Gamma_{\mathrm{max}}^{\mathrm{PL}} = 1.21 + 0.21_{\mathrm{stat}} + 0.1_{\mathrm{sys}} = 1.52$ and the BPL fit is not preferred over the PL fit ($P_{\mathrm{(BPL\;vs.\;PL)}} = 0.9488$).}
\label{Fig:Results1ES1101}
\end{figure}

To present the results for individual spectra in a compact and non-repetitive
way, we sort our preselected spectra into three categories. For each category
we select one prototype spectrum (following similar criteria to those in
Sect.~\ref{sec:tevspectra}), for which we give results. The
categories considered are:
\begin{description}
\item[\it "Nearby and well-measured".] As prototype spectrum we select the
Mkn\,501 spectrum  ($z = 0.034$) recorded by HEGRA during a major TeV flare in
1997. The relatively hard spectrum provides good statistics from
800\,GeV up to 25\,TeV. Other sources in this category are Mkn\,421 and
1ES\,1959+650. 1ES\,2344+514 and Mkn\,180 are at a comparable distance, but the
spectra do not have such high statistics. PKS\,2005-489 lies somewhere
between this and the following category. Spectra in this category mainly
provide limits in the FIR due to a pile-up at high energies.
\item[\it "Intermediate distance, wide energy range".] The prototype for this
category is the spectrum from H\,1426+428 at a redshift of $z = 0.129$, with
energies ranging from 700\,GeV up to 12\,TeV. PKS\,2155-304 is at a comparable
distance, and its measured spectrum has a better statistic, but the highest energy point is only at
2.5\,TeV; and the spectrum is much softer.
\item[\it "Distant source, hard spectrum".] The most distant TeV blazar
discovered so far with a published energy spectrum is 1ES\,1101-232 at a redshift of $z = 0.186$. Its spectrum
is hard and ranges from 160\,GeV to 3.3\,TeV. H\,2356-309 and 1ES\,1218+304 are
at similar distances, but the statistics are not as good and/or the spectrum is
softer. This makes 1ES\,1101-232  the natural choice for a prototype spectrum
in this category.
\end{description}

As the limit on the EBL for the individual spectra (and for the combined results in
Sect.~\ref{Section:CombinedResults} as well), we define the envelope shape
of all allowed EBL shapes. In most cases a single EBL shape represents the highest
allowed shape only for a small wavelength interval. Consequently the envelope
shape consists of a number of segments from different EBL shapes. One exception
occurs, if the maximum shape tested in the scan is allowed. In this case the
spectrum does not constrain the EBL density. In general, a spectrum
constrains the EBL density when the envelope shape lies below the maximum
shape tested in the scan.

Given the energy range of the spectra, the limit is
only valid for wavelengths $\lambda_{\mathrm{Lim}}$:
\begin{equation}
\label{eq:wrange}
 \lambda^{\mathrm{max}} (E_{min}) \; < \; \lambda_{\mathrm{Lim}} \; < \; \lambda^{\mathrm{thresh}} (E_{max}) 
 \end{equation}
where $\lambda^{\mathrm{max}} (E_{min})$ is the wavelength for which the cross
section for pair production with the lowest energy point of the VHE spectrum
$E_{min}$ is maximized (following Eqn.~(\ref{Formula:EBLRange})).
The variable $\lambda^{\mathrm{thresh}} (E_{max})$ is the threshold wavelength for pair
production with the highest energy point of the VHE spectrum $E_{max}$. This
wavelength range roughly reflects the sensitivity of the exclusion criteria.
There is, of course, some freedom of choice for the wavelength range, given the
complexity of the criteria.

In the following three sections the results for the individual prototype
spectra for the \textit{realistic} scan with $\Gamma_{\mathrm{max}} > 1.5$ are
summarized. The results for the \textit{extreme} scan with
$\Gamma_{\mathrm{max}} > 2/3$ for all prototype spectra are presented
in Sect.~\ref{Subsection:ResultsG0.6}.


\subsection{Nearby and well-measured: Mkn~501 (HEGRA)}\label{subsec:mkn501g1.5}

The envelope shape derived for the Mkn\,501 spectrum is shown in
Fig.~\ref{Fig:LimitMkn501} in comparison with the maximum and minimum EBL shapes
tested in the scan. Although 7\,766\,674 out of 8$\,$064$\,$000 EBL shapes
(96.3\%) can be excluded, the effective limit is only constraining in the 20
to 80$\,\mu$m wavelength region, where it lies below the maximum tested shape.
Note that our method is not only testing the overall level of the EBL density 
but is also sensitive to its structures. Despite the fact that, for the Mkn\,501 spectrum, almost all of the tested EBL shapes are excluded, certain types of shapes are allowed, independent of their respective EBL density level. This can be illustrated with an EBL shape, which 
has a power law dependency $n(\epsilon) \sim \epsilon^{-1}$ or $\nu I_{\nu} \sim \lambda^{-1}$. Then the optical depth is independent of the energy of VHE photons and the intrinsic TeV blazar spectrum has the same shape as the observed one \citep{aharonian:2001b}. With such a type of shape, an allowed high EBL density level in the MIR can be constructed by choosing a corresponding high density level in the optical/NIR.  

The rejection power in the FIR results mainly from the hard intrinsic
Mkn~501 spectrum above $\sim$5\,TeV that often can only be described by an
exponential or super exponential rise. Two of the limiting shapes, together with
the resulting intrinsic spectra and fit functions, are shown in
Fig.~\ref{Fig:ResultsMkn501} and the corresponding fit results can be found in
Table~\ref{Table:resmkn501}. In the table, the fit probabilities are 
given in the column ``$P_{\mathrm{Fit}}$''. The parameters are put in parentheses, if
the corresponding fit has a low probability or it is not preferred over 
a fit by a function with less free parameters (result of the likelihood ratio test, indicated in the column ``Pref''). The fit with a PL function did not meet our acceptance criteria, and the function is omitted from the figure for the sake of legibility. For the two shapes displayed here, 
the pile up at high energies is already visible but not yet significant.

\subsection{Intermediate distance, wide energy range: H1426+428}

Using the H\,1426+428 energy spectrum 5\,571\,772 EBL shapes (corresponding to
69.09\% of all shapes) are excluded. The resulting envelope shape is displayed
in Fig.~\ref{Fig:LimitH1426}, together with all highest allowed shapes. The
limit is constraining from $\sim$6 to $\sim$20$\,\mu$m and the constraints are
not very strong. This is illustrated in Fig.~\ref{Fig:ResultsH1426}, where two
representative highest allowed EBL shapes and the resulting intrinsic spectra
are shown, as are the relevant fit functions. Shape~A is almost as high as
the maximum shape of the scan. Although the intrinsic spectrum is already
convex, the relative low statistics of the spectrum even allows for an
acceptable fit by a simple PL and results in large errors on the fit parameters
for the BPL. Consequently the EBL shape cannot be excluded. Shape~B has a high
peak in the O/IR wavelength region but lies below all upper limits in the FIR.
The resulting intrinsic spectrum is best described with a simple PL with a
maximum slope $\Gamma_{\mathrm{max}}^{\mathrm{PL}} = 1.63 > 1.5$ and is
therefore allowed.

\subsection{Distant source, hard spectrum: 1ES~1101-232} 

The spectrum from 1ES\,1101-232 gives the strongest constraints for all
individual spectra, even though slightly less EBL shapes (7\,706\,625, 95.57\%
of all shapes) are excluded than in the case of Mkn\,501. This is due to the
different sensitivities in EBL wavelength; the 1ES\,1101-232 spectrum is only
sensitive up to ~10$\,\mu$m. The resulting maximum shapes and the envelope
shape are shown in Fig.~\ref{Fig:Limit1ES1101}. The envelope shape constraints
the EBL density in the wavelength range from $\sim$0.4 to $\sim$10$\,\mu$m and
clearly excludes the NIR excess claimed by \citet{matsumoto:2005a}
as being extragalactic. At EBL
wavelengths of $\sim$2$\,\mu$m the limit is consistent with the limit derived by
\citet{aharonian:2006:hess:ebl:nature} for the same source with a different
technique (see Sect.~\ref{Section:CombinedResults}).
Two representative highest allowed shapes and the corresponding intrinsic
spectra are shown in Fig.~\ref{Fig:Results1ES1101}. Shape A illustrates the
intrinsic spectrum for a high EBL density in the UV/O, while shape B is the
maximum shape for wavelengths around 3$\,\mu$m. For both shapes the examined
fit parameters are close to the allowed limits (see caption for values), as
expected for highest allowed shapes.

\subsection{\textit{Extreme} case: $\Gamma_{\mathrm{max}} > 2/3$}
\label{Subsection:ResultsG0.6}

\begin{figure*}[t,b]
\centering
\begin{minipage}[c]{0.45\textwidth}
\includegraphics[width=\textwidth]{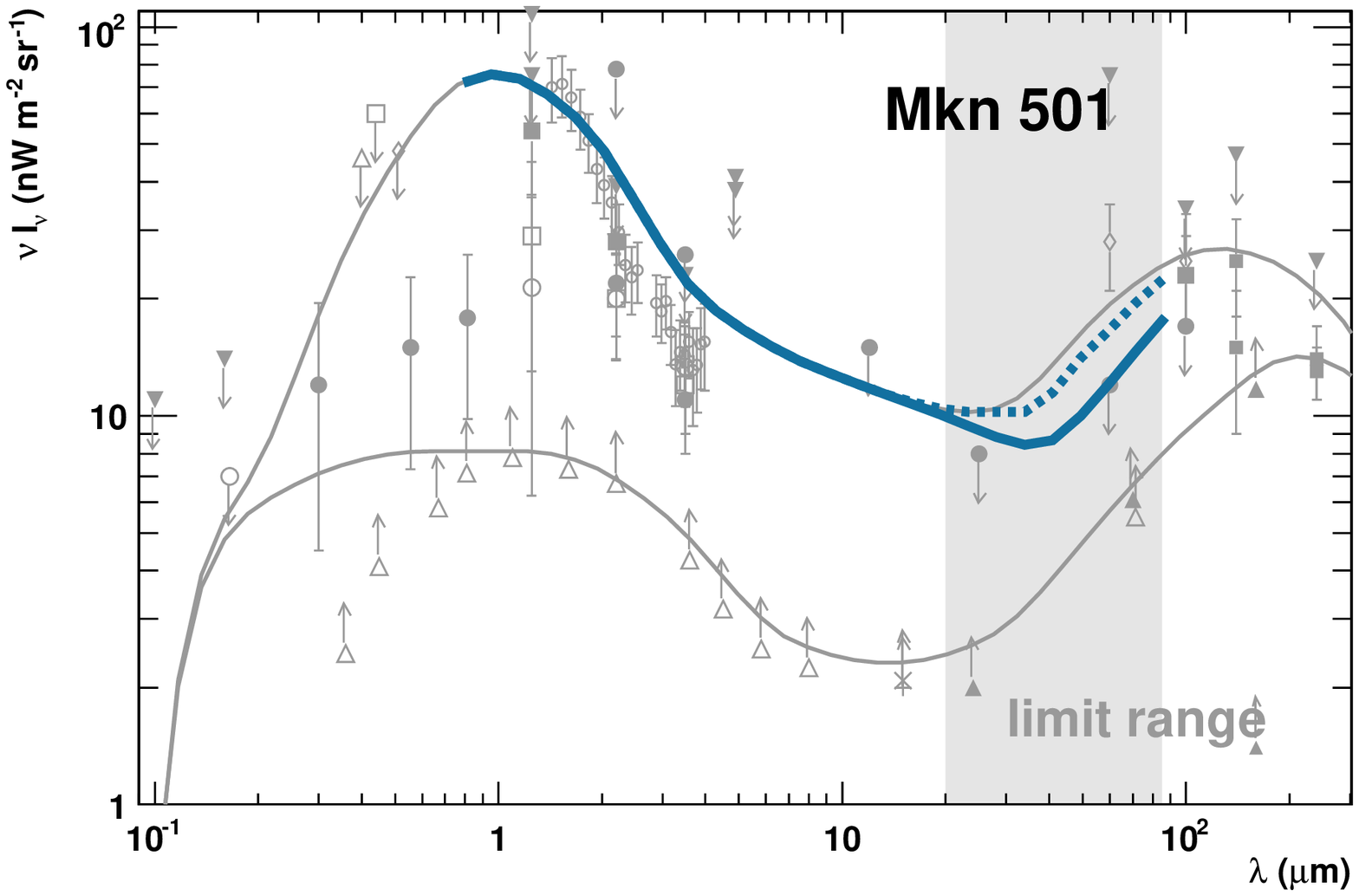}
\end{minipage}\hspace{0.9cm}%
\begin{minipage}[c]{0.45\textwidth}
\includegraphics[width=\textwidth]{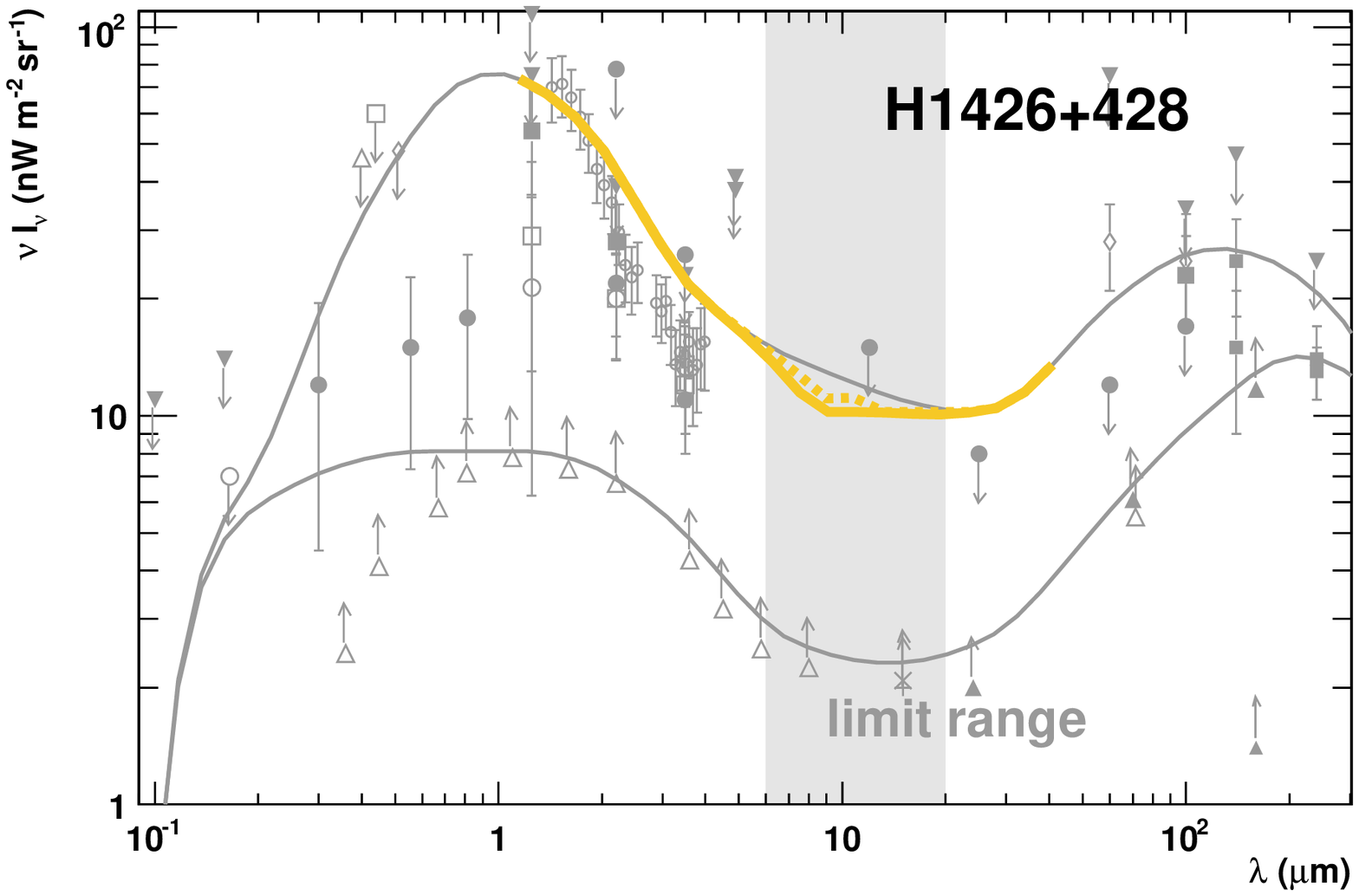}
\end{minipage} \\
\begin{minipage}[c]{0.45\textwidth}
\includegraphics[width=\textwidth]{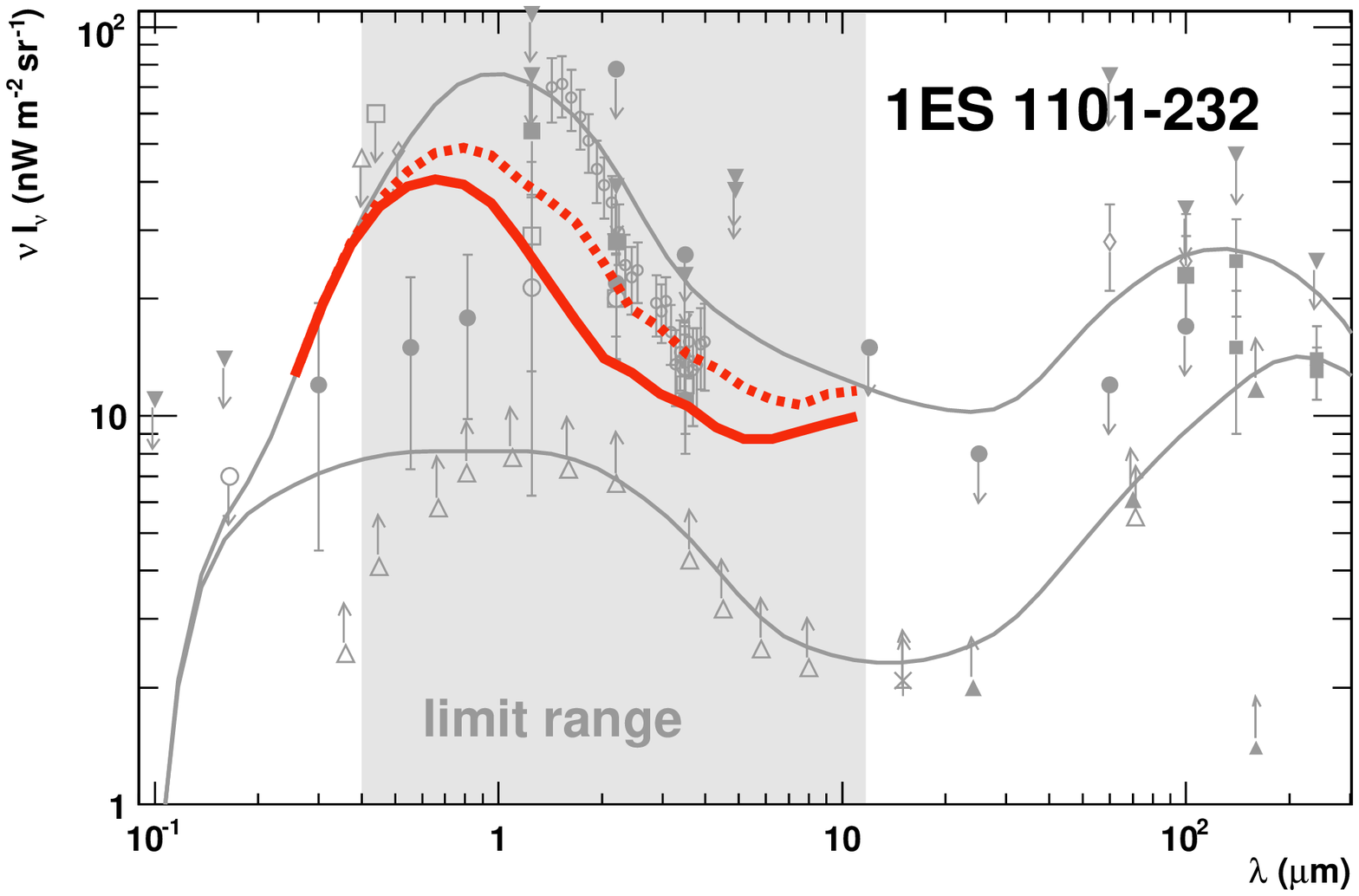}
\end{minipage}\hspace{0.9cm}%
\begin{minipage}[c]{0.45\textwidth}
\caption{Result for individual spectra, comparing the \textit{realistic} case with $\Gamma_{\mathrm{max}} > 1.5$ (thick colored solid line) with the \textit{extreme} case with $\Gamma_{\mathrm{max}} > 2/3$ (thick colored dashed line), overlaid on the EBL measurements and the minimum and maximum shapes of the scan (grey markers and lines). The grey boxes indicate the validity range of the limit for the \textit{realistic} scan as in Figs.~\ref{Fig:LimitMkn501}, \ref{Fig:LimitH1426}, and \ref{Fig:Limit1ES1101}. Note that the validity range of the limits from the \textit{extreme} scan are quite similar. \textit{Upper Left Panel:} Results for Mkn~501 (HEGRA). \textit{Upper Right Panel:} Results for H1426+428. \textit{Lower Left Panel:} Results for 1ES~1101-232.}
\label{Fig:ResultSpectraG0.6}
\end{minipage}
\end{figure*}
 
The upper limits for Mkn\,501, H1426+428 and 1ES\,1101-232 for the \textit{extreme} case with $\Gamma_{\mathrm{max}} > 2/3$ are shown in Fig.~\ref{Fig:ResultSpectraG0.6} in comparison to the limits derived for the \textit{realistic} case. In the case of Mkn\,501, a similar number of EBL shapes (94.13\%) as in the \textit{realistic} scan are excluded, but the effective limit is less constraining (Fig.~\ref{Fig:ResultSpectraG0.6}, Upper Left Panel). For H1426+428 the limit almost remains at the same level, still close to the maximum shape tested in the scan (Fig.~\ref{Fig:ResultSpectraG0.6}, Upper Right Panel). For 1ES\,1101-232 the limit in the UV/O lies a factor of 1.2 to 1.8 higher than the limit in the \textit{realistic} case (Fig.~\ref{Fig:ResultSpectraG0.6}, Lower Left Panel). In the wavelength region from 2 to 4$\mu$m, the NIR excess claimed by \cite{matsumoto:2005a} is now compatible with the limit. In the 1 to 2$\mu$ region the limit still lies clearly below the claimed NIR excess.


\section{Combined results}
\label{Section:CombinedResults}

\begin{table}
\begin{center}
\caption{Number of excluded EBL shapes for the {\it realistic} scan for individual spectra (column two) as well as the number of the intrinsic spectra, which could not be fitted satisfactorily (column three).$^{\mathrm a}$}
\begin{tabular}{lcc}
\hline \hline
Spectrum & \#Shapes Excluded & \#Shapes No Fit \\
\hline
Mkn 421 (MAGIC) &   1\,756\,869 (21.79\%) &   59\,258 (0.94\%)  \\
Mkn 421 (HEGRA) &   888\,575 (11.02\%) & 0  \\
Mkn 421 (Whipple) &  2\,287\,059 (28.36\%)  & 0 \\
Mkn 501 &  7\,760\,733 (96.24\%) &  1296 (0.43\%)  \\
1ES\,2344+514 & 0 & 0 \\
Mkn\,180 & 0 & 0 \\
1ES\,1959+650 &  1\,086\,314 (13.47\%)  & 423 (0.01\%) \\
PKS\,2005-489 &  0 & 0 \\
PKS\,2155-304 &  4\,248\,872 (52.69\%)  & 0 \\
H\,1426+428 & 5\,571\,771 (69.09\%) &  2 (0.00\%)  \\
H\,2356-309 & 4\,657\,817 (57.76\%) & 0 \\
1ES\,1218+304 & 19\,540 (00.24\%) & 0 \\
1ES\,1101-232 & 7\,706\,624 (95.57\%)  & 0  \\
\hline
\end{tabular} \label{Table:ShapeExStat}
\end{center}
{\footnotesize
$^{\mathrm a}$ The percentage of these numbers to the number of all shapes (8\,064\,000) for column two and to the number of the allowed shapes for this spectrum for column three is also quoted. If the number of shapes is zero, the percentage is omitted.}
\end{table}

\begin{figure}[t,b]
\centering
\includegraphics[width=0.45\textwidth]{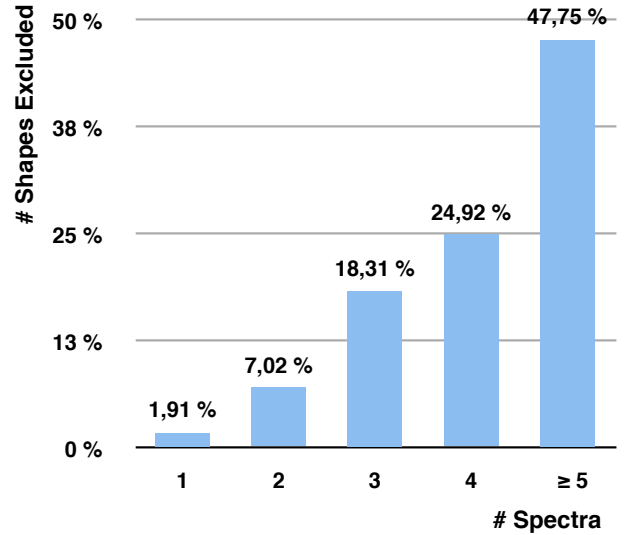}
\caption{Number of excluded shapes as percentage of the total number of shapes vs. the number of spectra, that excluded the shape. A large fraction of shapes (98.08\%) is excluded by more than one spectra.}
\label{Fig:NShapeExVsNSpec}
\end{figure}
 
\begin{figure*}[t,b]
\centering
\begin{minipage}[c]{0.6\textwidth}
\includegraphics[width=\textwidth]{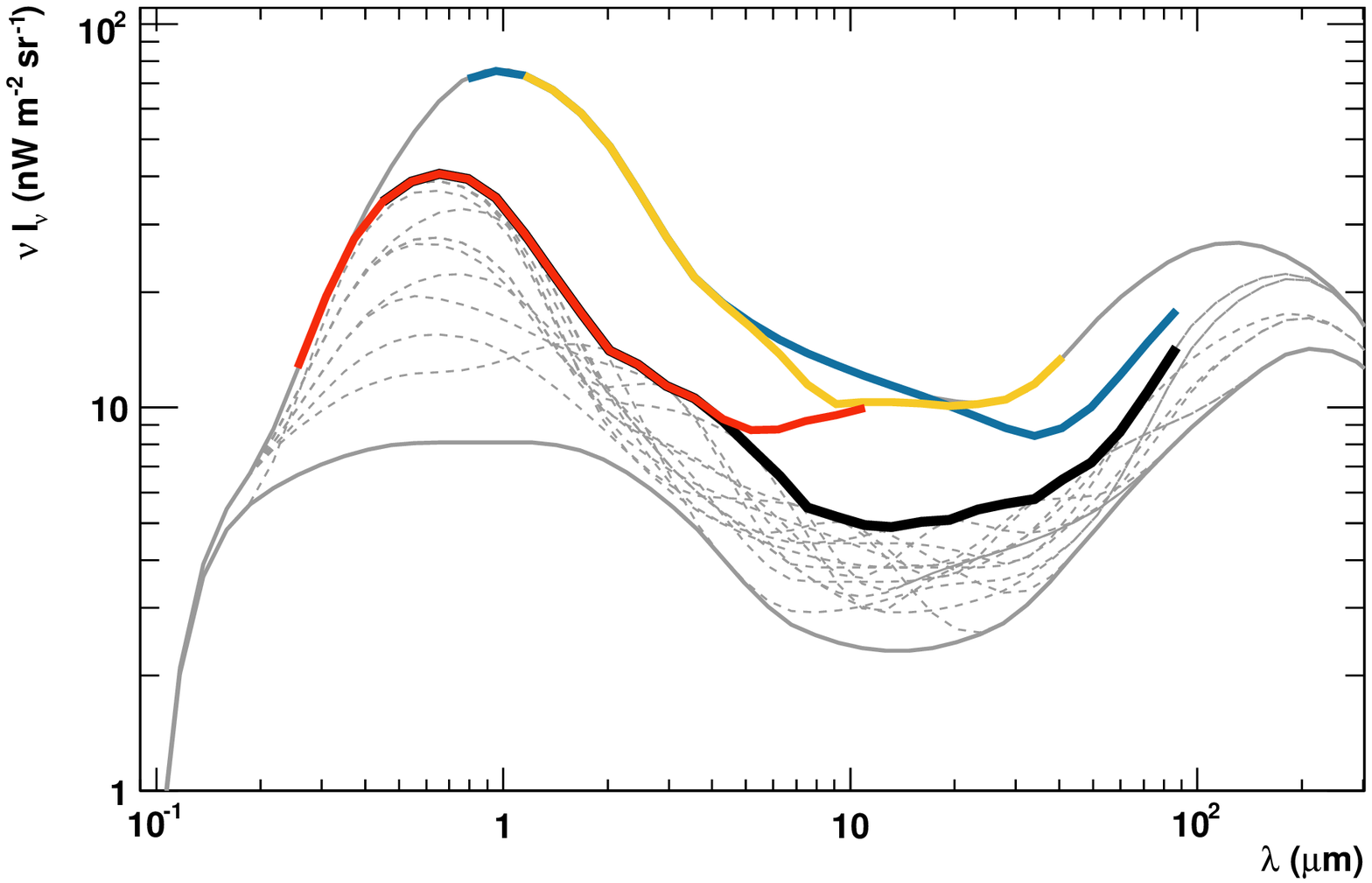}\\
\includegraphics[width=\textwidth]{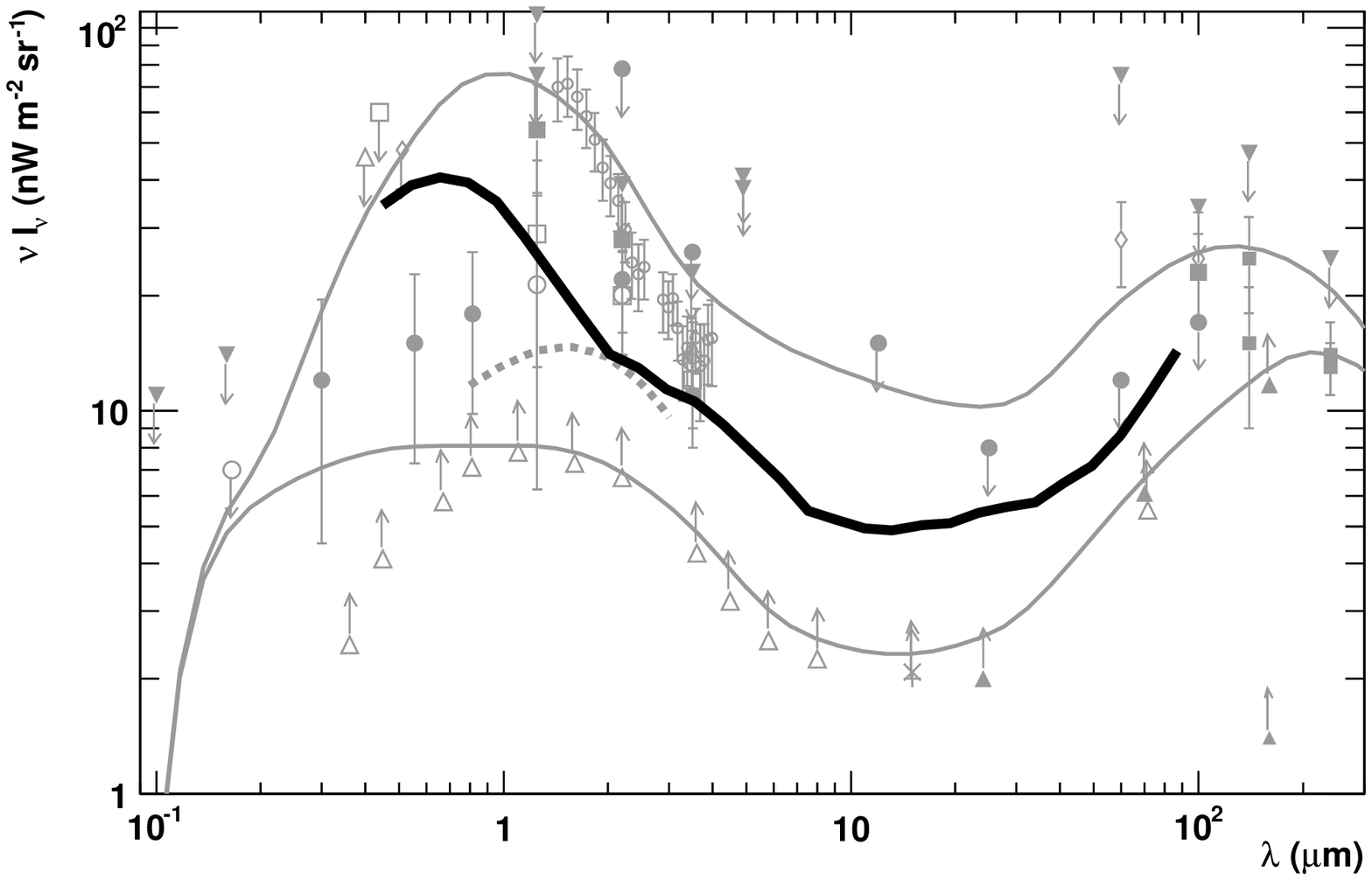}
\end{minipage}\hspace{0.5cm}%
\begin{minipage}[c]{0.25\textwidth}
\caption{Combined results for the \textit{realistic} scan. \textit{Upper Panel:} All highest allowed shapes of the combined scan (dashed grey lines) and the corresponding envelope shape (solid black line) in comparison to the limits for individual spectra: Mkn\,501 (solid blue line), H\,1426+428 (solid yellow line), and 1ES\,1101-232 (solid red line). The minimum and maximum shapes of the scan are also shown (grey lines). \textit{Lower Panel:} The combined limit from the \textit{realistic} scan (solid black line) in comparison to direct measurements and limits (grey marker). The grey dashed curve around 2$\,\mu$m is the limit derived by \cite{aharonian:2006:hess:ebl:nature}.}
\label{Fig:CombinedResults01}
\end{minipage}
\end{figure*}

\begin{figure*}[t,b]
\centering
\begin{minipage}[c]{0.6\textwidth}
\includegraphics[width=\textwidth]{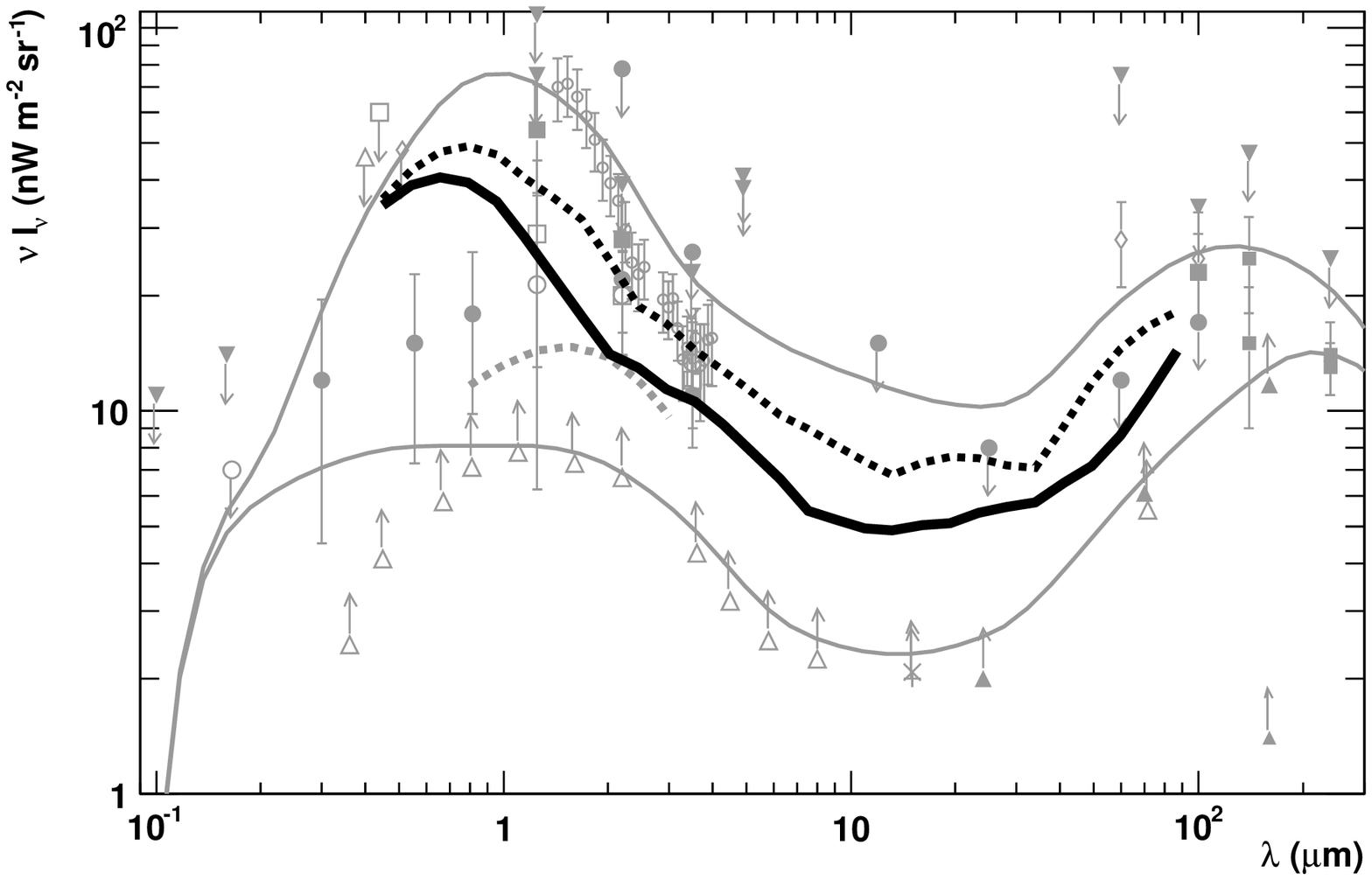}
\end{minipage}\hspace{0.5cm}%
\begin{minipage}[c]{0.25\textwidth}
\caption{Combined results from the \textit{extreme} scan (dashed black line) in comparison to the result from the \textit{realistic} scan (solid black line). Grey lines are the minimum and the maximum shapes tested in the scan. Grey markers are direct measurements and limits.}
\label{Fig:CombinedResultsExtremeScan}
\end{minipage}
\end{figure*}
 
 The number of excluded shapes for all individual spectra for the \textit{realistic} scan are summarized in Table~\ref{Table:ShapeExStat}. One finds that some spectra (namely 1ES\,2344+514, Mkn\,180, and PKS\,2005-489) excluded none of the EBL shapes of the scan. This mainly is a result of the low statistics in the spectra, hence the large errors on the fit parameters. The highest number of excluded shapes and the strongest constraint in the scan result from the previously presented prototype spectra Mkn\,501, H\,1426+428, and 1ES\,1101-232.

We now combine the results from all spectra by excluding all shapes of the scan, which are excluded by at least one of the spectra. First we present results for the \textit{realistic} scan. A large fraction of the shapes (98.08\%) are excluded by more than one spectra (see Fig.~\ref{Fig:NShapeExVsNSpec}), which strengthens our results. In total, we exclude 8\,056\,718 shapes, which leaves us with only 7\,282 (0.01\%) allowed shapes. The resulting maximum shapes (dashed lines) and the envelope shape for the combined results are shown in the upper panel of Fig.~\ref{Fig:CombinedResults01} in comparison to the results from the individual prototype spectra. In the optical-to-near-infrared the combined limit follows exactly the limit derived from the 1ES\,1101-232 spectrum. For longer wavelengths, however, the combined limit lies significantly below the limits derived from individual spectra. This is particularly striking in the MIR, where the H\,1426+428 spectrum provided only weak limits and the combined limit now gives considerable constraints. This can be understood from the fact that, for a high EBL density in the MIR, a high density in the optical to NIR is needed, so that the spectra (Mkn\,501, H\,1426+428) will not get too hard. These high densities in the optical-to-NIR are now excluded by the 1ES\,1110-232 spectrum, which therefore result in stronger constraints in the MIR.

The combined limit for the \textit{realistic} scan in comparison to the direct
measurements and limits is shown in the lower panel of Fig.~\ref{Fig:CombinedResults01}. In the optical-to-NIR one finds that the combined limit lies
significantly below the claimed NIR excess by \citet{matsumoto:2005a}. It is
compatible with the detections reported by \citet{dwek:1998b},
\citet{gorjian:2000a}, \citet{wright:2000a}, and \citet{cambresy:2001a}. In the
same figure the limit reported by the H.E.S.S. collaboration for 1ES\,1101-232
derived with a different technique \citep{aharonian:2006:hess:ebl:nature} is
shown and at wavelengths around 2 - 3$\,\mu$m
it is in good agreement with the limit derived here; but for shorter
wavelengths our limit lies significantly higher. While for the H.E.S.S. limit
comparable exclusion criteria for the spectra were used,
a fixed reference shape scaled in the level of the EBL density was used to calculate the EBL limit%
\footnote{In addition to the fixed shape, several other possible 
EBL components, like a bump in the UV, were examined \citep{aharonian:2006:hess:ebl:nature}.},
which presumably causes the differences at smaller wavelengths. 
In the MIR to FIR our combined limit lies
below all previously reported upper limits from direct measurements and
fluctuation analysis. For EBL wavelengths greater than 2$\,\mu$m, it is only
about a factor of 2 to 2.5 higher than the absolute lower limit from source
counts, leaving very little room for additional contributions to the EBL in
this wavelength region. In the FIR it lies more than a factor of $\sim2$ below
the claimed detection by \citet{finkbeiner:2000a}.

In the \textit{realistic} scan, after combining the results from Mkn\,501,
H\,1426+428, and 1ES\,1110-232, adding more spectra does not further strengthen the limit (though marginal more shapes are excluded). Hence, for the \textit{extreme} scan, we
only combine the results from these three spectra. The limit for the \textit{extreme}
scan is shown in Fig.~\ref{Fig:CombinedResultsExtremeScan} in comparison to the limit from the
\textit{realisistic} scan.
The limit from the \textit{extreme} scan lies for all EBL wavelengths above the
limits from the realistic scan and is a factor of 2.5 to 3.5 higher than the
lower limits from source counts. In the optical-to-NIR the combined limit
again follows the limit derived from the 1ES\,1101-232 spectrum. As stated for
the individual spectrum the NIR excess claimed by \cite{matsumoto:2005a} for
wavelengths above 2$\,\mu$m is now compatible with the limit, but for
shorter wavelengths it is still clearly excluded.
 
\begin{figure*}[t,b]
\centering
\begin{minipage}[c]{0.6\textwidth}
\includegraphics[width=\textwidth]{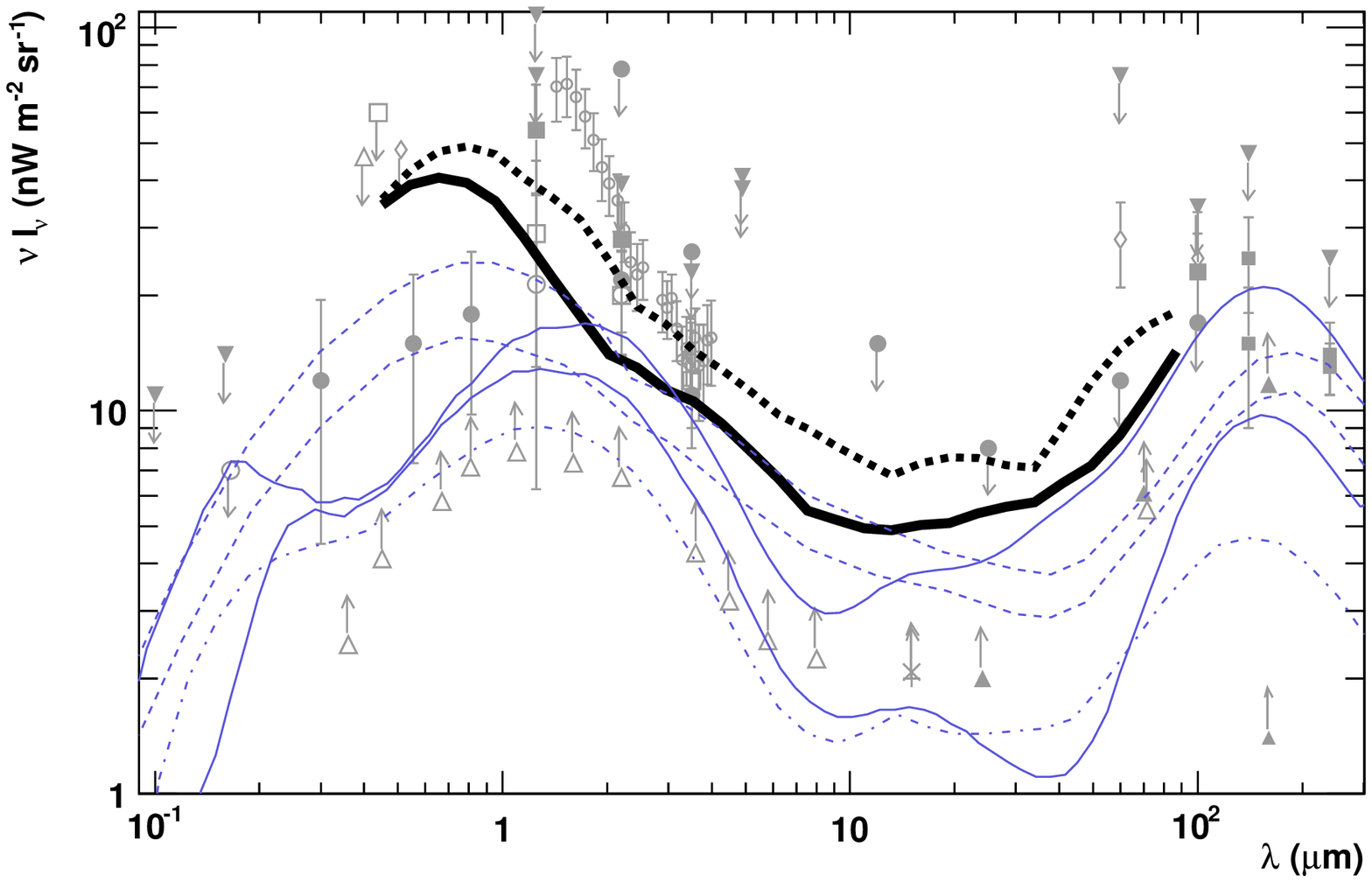}
\end{minipage}\hspace{0.5cm}%
\begin{minipage}[c]{0.25\textwidth}
\caption{EBL limits from the \textit{realistic} (thick solid black curve) and the \textit{extreme} scan (thick dashed black curve) in comparison to different EBL models at $z = 0$ (blue curves): updated version of the \citet{kneiske:2002a} high and low models (solid), \citet{stecker:2006a} fast and baseline evolution models (dashed), \citet{primack:2005a} (dashed dotted).}
\label{Fig:LimitVsModels}
\end{minipage}
\end{figure*}

In Fig.~\ref{Fig:LimitVsModels} the limits derived in this paper are shown in comparison to predictions from EBL models by \citet{kneiske:2002a}, \citet{stecker:2006a}, and \citet{primack:2005a} (at $z = 0$).
Most models lie below our \textit{realisitc} limits. Only the high model of \citet{kneiske:2002a} is slightly higher in the NIR, and the fast evolution model of \citet{stecker:2006a} is marginally higher in the MIR.  These two models are also excluded when applying the criteria from Sect.~\ref{sec:criteria}, while the other models are allowed. Noteworthy is that the \citet{primack:2005a} and the low model from \citet{kneiske:2002a} are below the lower limits derived from galaxy counts in the MIR and FIR.

\section{Summary and conclusions} \label{Section:Conclusion}

In this paper we present a new method of constraining the density of the
extragalactic background light (EBL) in the optical-to-far-infrared wavelength
regime using VHE spectra of TeV blazars. We derive strong upper limits, which
are only a factor 2 to 3 higher than the lower limits determined from source counts.
Unlike what has been done in many previous studies of this kind, we do not
use one or a few pre-defined EBL shapes or models, but rather a scan on
a grid in EBL density vs. wavelength. This grid covers wavelengths from
0.1\,$\mu$m to 1000\,$\mu$m and spans in EBL density from the lower limits from
source counts to the upper limits determined from direct measurements and
fluctuation analyses. By iterating over all scan points, we test a large set of
different EBL shapes (8$\,$064$\,$000 in total). Each EBL shape is described by
a  spline function, which allows us to calculate the optical depth of VHE
$\gamma$-rays for this shape via a simple summation instead of solving three
integrals numerically.  The resolution of the scan for sharp bumps or dips in
the EBL is limited by the choice of the grid and the order of the spline.

To test these EBL shapes, we use the measured VHE $\gamma$-ray spectra from
distant sources, which carry an imprint of the EBL from attenuation via pair
production. We included spectra from all known TeV blazars (13 in total),
making this the most complete study of this type up to now. All spectra are
scrutinized with the same robust algorithm. For each spectrum and each EBL
shape, the intrinsic spectrum is calculated and an analytical description of the
intrinsic spectrum is determined by fitting several functions. The fit
parameters and errors of the functions are subsequently used to evaluate
whether the intrinsic spectrum is physically feasible. We use two conservative
criteria from the theory of VHE $\gamma$-ray emission in TeV blazars. The EBL
shape is excluded if (i) a part of the intrinsic spectrum is harder than a
theoretical limit or (ii) the intrinsic spectrum shows a significant pile-up at
high energies. Since there is some spread in the predictions from theory, we
included two independent scans for two theoretical limits on the hardness of
the spectrum: (1) the \textit{realistic} case for a photon index of $\Gamma >
1.5$ and (2) the \textit{extreme} case with $\Gamma > 2/3$.

We present limits derived from individual VHE $\gamma$-ray spectra. The
strongest constraints result from the 1ES\,1101-232 spectrum in the wavelength
range from 0.8 and 10\,$\mu$m, mainly due to the hard spectrum and its large
distance to earth. An extragalactic origin of the claimed excess in the NIR
between 1 and 2$\,\mu$m \citep{matsumoto:2005a} can be excluded even by the
\textit{extreme} scan. This result confirms the results from the recent
publication of \citet{aharonian:2006:hess:ebl:nature}. In the FIR (20 and
80\,$\mu$m), strong upper limits are provided by the nearby TeV blazar
Mkn\,501. H\,1426+428, situated at an intermediate distance, provides  some
constraints on the EBL density in the MIR (between 1 and 15\,$\mu$m),
connecting the upper limits derived from 1ES\,1101-232 and Mkn\,501. Most of
the other spectra provide constraints in certain wavelength ranges, but they
are not as strong as the limits from the previously described sources.

By combining the results from all spectra, we find that the upper limits become
much stronger compared to the limits derived for individual spectra, especially
in the wavelength range between 4 and 60\,$\mu$m. In the wavelength region from 2 to 80$\,\mu$m, the combined
limit for the \textit{realistic} case lies below the upper limits derived from
fluctuation analyses of the direct measurements \citep{kashlinsky:2000a} and is
just a factor of 2 to 2.5 above the absolute lower limits from source counts.
This makes it the most constraining limit in the MIR to FIR region so far.  As
expected, the upper limits from the \textit{extreme} scan are less constraining
(factor of 2.5 to 3.5 above the lower limit from source counts), but still substantial over this wide wavelength range.

The derived upper limits can be interpreted in two different ways: 
\begin{enumerate}
 \item The EBL density in the optical-to-FIR is significantly lower than suggested by direct
measurements, and the actual EBL level seems to be close to the existing lower limits. 
It can thus be concluded that experiments like HST, ISO, and Spitzer resolve
most of the sources in the universe. This would indicate that there is very little room left
for a significant contribution of heavy and bright Population III stars to the EBL density in
this wavelength region.
 \item The assumptions used for this study are not correct.  This would require
a revision of the current understanding of TeV blazar physics and models, which
has so far been fairly successful in modeling multi-wavelength data from the radio to
the VHE for all detected sources.  
\end{enumerate}

We estimate that the main contribution to the systematic errors arises
from the minimum width of the EBL structures that can be resolved by the
scan.  Although the choice of the grid spacing, which defines the minimum
width of the EBL structures, is physically well motivated, there are
arguments that even thinner structures could be realized in nature due to, say
absorption effects. We tested a 20\% smaller grid spacing, resulting in much
less realistic, but still possible, EBL structures, with the 1ES\,1101-232
spectrum. The resulting limits on the EBL are 20 to 30\% higher than the ones
presented in Section~5.3. We therefore conclude that the overall
contribution from the grid setup to the systematic error is at most 30\%.

Another contribution to the systematic error originates from 
not considering the evolution of the EBL in our method.
To estimate the effect of the evolution of the EBL for the most
distant sources, we calculate the late contribution to the EBL in the redshift 
interval between $z = 0$ and $z = 0.2$. Here, we utilize the EBL
model by \citet{kneiske:2002a} (updated version is used, Kneiske private
communication). We obtain a wavelength dependent contribution, which
has a maximum value of 3-4\,nW\,m$^{-2}$sr$^{-1}$ in the optical to NIR
(the relevant wavelength region for the distant sources in our scan). Given
 that these values are derived with the extreme assumption that all 
 late emission occured instantaneously at z=0, we estimate the systematic 
 error arising from the EBL evolution to 10\% of the derived limits. 
 Note that \citet{aharonian:2006:hess:ebl:nature} estimated an error of $<$10\% for 
 the most distant TeV blazar in our study 1ES\,1101-232.

To study the effect of the uncertainties in the absolute energy scale of IACT measurements, we calculate a limit on the EBL utilizing the energy spectra from Mkn\,501, H\,1426+428, and 1ES\,1101-232 shifted by 15\% to lower energies (for $\Gamma > 1.5$, \textit{realistic} scan). Since the optical depth is increasing with energy, the energy shift reduces the attenuation of the spectra, especially reducing the effect of a possible pile-up at the highest energies. The difference between the limit derived from the energy shifted spectra and the \textit{realistic} limit is negligible in the optical to NIR ($<$6\,$\mu$m), while there is a moderate effect in the MIR to FIR on the level of 1\,--\,4.5\,nW\,m$^{-2}$sr$^{-1}$. We therefore conclude that the systematic error in the MIR to FIR ($\lambda > 6\,\mu$m) introduced by the uncertainties in the absolute energy scale of IACT measurements is close to 10--45\% of our \textit{realistic} limit.

Adding the individual errors quadratically, we finally conclude that the (very conservative) systematic error on the upper limit is about 32\% in the optical-to-NIR and 33\,--\,55\% in MIR to FIR.

The strong constraints derived in this paper only allow for a low level of the
EBL in the optical to the far-infrared, suggesting that the universe is more
transparent to VHE $\gamma$-rays than previously thought. Hence, we expect
detections of many new extragalactic VHE sources in the next few years. Further
multi-wavelength studies of TeV blazars will improve the understanding of the
underlying physics, which will help to refine the exclusion criteria for the
VHE spectra in this kind of study. The upcoming GLAST satellite experiment,
operating in an energy range from 0.1 up to $\sim$100 GeV, will allow
such studies of the EBL to be extended to the ultraviolet-to-optical wavelength region. A
detailed analysis of the impact of our constraints on the contribution from
Population III stars to the EBL is in preparation.


\begin{acknowledgements}
The authors would like to thank J.~Ripken for fruitful discussion of
splines and the careful reading of the manuscript.
The authors also wish to
thank M.~Beilicke, R.~Cornils, F.~Goebel, N.~Goetting, and G.~Heinzelmann for stimulating comments and help in preparing this manuscript.
The authors also would like to thank the anonymous referee for the helpful comments, which improved the manuscript.
D.~M. acknowledges the financial support of  the MPI f\"ur Physik, Munich.
M.~R. acknowledges the financial support of the University of Hamburg and the BMBF.
This research made use of NASA's Astrophysics Data System.
\end{acknowledgements}

\appendix

\section{Likelihood Ratio Test}\label{Appendix:Likelihood}

The likelihood ratio test \citep{eadie:1998a} is a standard statistical tool to test
between two hypotheses whether an improvement to a fit quality (quantified by
corresponding $\chi^2$ values) is expected from a normal distribution or if it is
significant.  By fitting two functional forms to the intrinsic spectrum,  one
obtains values of the likelihood functions $L_A$ and $L_B$. If hypothesis A is
true, the likelihood ratio $R = -\ln(L_A/L_B)$ is approximately $\chi^2$
distributed with $N$ degrees of freedom.  $N$ is the difference between numbers
of degrees of freedom of hypothesis A and hypothesis B.  One defines a
probability \begin{equation} \label{eq:probab} P =
\int\limits_{0}^{R_{\mathrm{meas}}} p\,(\chi^2) \, \mathrm{d\chi^2}
\end{equation} where $p\,(\chi^2)$ is the $\chi^2$ probability density function
and $R_{meas}$ the measured value of $R$. Hypothesis A will be rejected (and
hypothesis B will be accepted) if $P$ is greater than the confidence level,
which is set to 95\%.


\def\Journal#1#2#3#4{{#4}, {#1}, {#2}, #3}
\def\NAT{Nature}
\def\AAA{A\&A}
\def\ApJ{ApJ}
\def\AJ{Astronom. Journal}
\def\Aph{Astropart. Phys.}
\def\ApJS{ApJSS}
\def\MNRAS{MNRAS}
\def\NIM{Nucl. Instrum. Methods}
\def\NIMA{Nucl. Instrum. Methods A}

\bibliographystyle{aa}
\bibliography{ms}

\begin{thebibliography}{81}
\expandafter\ifx\csname natexlab\endcsname\relax\def\natexlab#1{#1}\fi

\bibitem[{{Aharonian} {et~al.}(2005{\natexlab{a}}){Aharonian}, {Akhperjanian},
  {Aye}, {et~al.}}]{Aharonian2005:HESS.PKS2005}
{Aharonian}, F., {Akhperjanian}, A.~G., {Aye}, K.-M., {et~al.}
  2005{\natexlab{a}}, \aap, 436, L17

\bibitem[{{Aharonian} {et~al.}(2005{\natexlab{b}}){Aharonian}, {Akhperjanian},
  {Bazer-Bachi}, {et~al.}}]{Aharonian2005:HESS.PKS2155.MWL}
{Aharonian}, F., {Akhperjanian}, A.~G., {Bazer-Bachi}, A.~R., {et~al.}
  2005{\natexlab{b}}, \aap, 442, 895

\bibitem[{{Aharonian} {et~al.}(2006{\natexlab{a}}){Aharonian}, {Akhperjanian},
  {Bazer-Bachi}, {et~al.}}]{aharonian:2006:hess:ebl:nature}
{Aharonian}, F., {Akhperjanian}, A.~G., {Bazer-Bachi}, A.~R., {et~al.}
  2006{\natexlab{a}}, \nat, 440, 1018

\bibitem[{{Aharonian} {et~al.}(2006{\natexlab{b}}){Aharonian}, {Akhperjanian},
  {Bazer-Bachi}, {et~al.}}]{Aharonian2006:HESS.H2356}
{Aharonian}, F., {Akhperjanian}, A.~G., {Bazer-Bachi}, A.~R., {et~al.}
  2006{\natexlab{b}}, \aap, 455, 461

\bibitem[{{Aharonian} {et~al.}(2006{\natexlab{c}}){Aharonian}, {Akhperjanian},
  {Bazer-Bachi}, {et~al.}}]{aharonian2006:hess:pg1553}
{Aharonian}, F., {Akhperjanian}, A.~G., {Bazer-Bachi}, A.~R., {et~al.}
  2006{\natexlab{c}}, \aap, 448, L19

\bibitem[{{Aharonian} {et~al.}(2006{\natexlab{d}}){Aharonian}, {Akhperjanian},
  {Bazer-Bachi}, {et~al.}}]{aharonian:2006:hess:m87:science}
{Aharonian}, F., {Akhperjanian}, A.~G., {Bazer-Bachi}, A.~R., {et~al.}
  2006{\natexlab{d}}, Science, 314, 1424

\bibitem[{{Aharonian}(2001)}]{aharonian:2001b}
{Aharonian}, F.~A. 2001, in Proceedings of 27th ICRC, Highlight papers,
  Hamburg, ed. R.~{Schlickeiser}, 250

\bibitem[{{Aharonian} {et~al.}(2002{\natexlab{a}}){Aharonian}, {Akhperjanian},
  {Beilicke}, {et~al.}}]{aharonian:2002e}
{Aharonian}, F.~A., {Akhperjanian}, A., {Beilicke}, M., {et~al.}
  2002{\natexlab{a}}, Astronomy and Astrophysics, 393, 89

\bibitem[{{Aharonian} {et~al.}(1999){Aharonian}, {Akhperjanian}, {Barrio},
  {et~al.}}]{aharonian:1999b}
{Aharonian}, F.~A., {Akhperjanian}, A.~G., {Barrio}, J.~A., {et~al.} 1999,
  Astronomy and Astrophysics, 349, 11

\bibitem[{{Aharonian} {et~al.}(2002{\natexlab{b}}){Aharonian}, Akhperjanian,
  Barrio, {et~al.}}]{aharonian:2002a}
{Aharonian}, F.~A., Akhperjanian, A.~G., Barrio, J.~A., {et~al.}
  2002{\natexlab{b}}, Astronomy and Astrophysics, 384, L23

\bibitem[{{Aharonian} {et~al.}(2003{\natexlab{a}}){Aharonian}, Akhperjanian,
  Beilicke, {et~al.}}]{aharonian:2003c}
{Aharonian}, F.~A., Akhperjanian, A.~G., Beilicke, M., {et~al.}
  2003{\natexlab{a}}, Astronomy and Astrophysics, 406, L9

\bibitem[{{Aharonian} {et~al.}(2003{\natexlab{b}}){Aharonian}, Akhperjanian,
  Beilicke, {et~al.}}]{aharonian:2003b}
{Aharonian}, F.~A., Akhperjanian, A.~G., Beilicke, M., {et~al.}
  2003{\natexlab{b}}, Astronomy and Astrophysics, 403, L1

\bibitem[{{Aharonian} {et~al.}(2003{\natexlab{c}}){Aharonian}, Akhperjanian,
  Beilicke, {et~al.}}]{aharonian:2003a}
{Aharonian}, F.~A., Akhperjanian, A.~G., Beilicke, M., {et~al.}
  2003{\natexlab{c}}, Astronomy and Astrophysics, 403, 523, astro-ph/0301437

\bibitem[{{Aharonian} {et~al.}(2002{\natexlab{c}}){Aharonian}, {Timokhin}, \&
  {Plyasheshnikov}}]{aharonian:timokhin:2001a}
{Aharonian}, F.~A., {Timokhin}, A.~N., \& {Plyasheshnikov}, A.~V.
  2002{\natexlab{c}}, \aap, 384, 834


\bibitem[{{Albert} {et~al.}(2006{\natexlab{a}}){Albert}, {Aliu}, {Anderhub},
  {et~al.}}]{albert:2006:magic:1ES1218}
{Albert}, J., {Aliu}, E., {Anderhub}, H., {et~al.} 2006{\natexlab{a}}, \apjl,
  642, L119

\bibitem[{{Albert} {et~al.}(2006{\natexlab{b}}){Albert}, {Aliu}, {Anderhub},
  {et~al.}}]{albert:2006:magic:mkn180}
{Albert}, J., {Aliu}, E., {Anderhub}, H., {et~al.} 2006{\natexlab{b}}, \apjl,
  648, L105

\bibitem[{{Albert} {et~al.}(2007{\natexlab{a}})}]{albert:2006:magic:pg1553}
{Albert}, J., {Aliu}, E., {Anderhub}, H., {et~al.} 
2007{\natexlab{a}},  \apjl, 654, L119

\bibitem[{{Albert} {et~al.}(2007{\natexlab{b}}){Albert}, {Aliu}, {Anderhub},
  {et~al.}}]{albert:2006:magic:mkn421}
{Albert}, J., {Aliu}, E., {Anderhub}, H., {et~al.} 
2007{\natexlab{b}}, ApJ, 663, 125

\bibitem[{{Bernstein} {et~al.}(2002){Bernstein}, {Freedman}, \&
  {Madore}}]{bernstein:2002a}
{Bernstein}, R.~A., {Freedman}, W.~L., \& {Madore}, B.~F. 2002, Astrononmy and
  Astrophysics, 571, 56

\bibitem[{{Bernstein} {et~al.}(2005){Bernstein}, {Freedman}, \&
  {Madore}}]{bernstein:2005a}
{Bernstein}, R.~A., {Freedman}, W.~L., \& {Madore}, B.~F. 2005, \apj, 632, 713

\bibitem[{{Brown} {et~al.}(2000){Brown}, {Kimble}, {Ferguson}, {Gardner},
  {Collins}, \& {Hill}}]{brown:2000a}
{Brown}, T.~M., {Kimble}, R.~A., {Ferguson}, H.~C., {et~al.} 2000, \aj, 120,
  1153

\bibitem[{{Cambr{\' e}sy} {et~al.}(2001){Cambr{\' e}sy}, {Reach}, {Beichman},
  \& {Jarrett}}]{cambresy:2001a}
{Cambr{\' e}sy}, L., {Reach}, W.~T., {Beichman}, C.~A., \& {Jarrett}, T.~H.
  2001, The Astrophysical Journal, 555, 563

\bibitem[{{Cortina} {et~al.}(2005)}]{cortina:2005a}
{Cortina}, J. {et~al.} 2005, in International Cosmic Ray Conference, Vol.~5,
  359

\bibitem[{{Costamante} {et~al.}(2003){Costamante}, {Aharonian}, {Ghisellini},
  \& {Horns}}]{costamante:2003a}
{Costamante}, L., {Aharonian}, F., {Ghisellini}, G., \& {Horns}, D. 2003, New
  Astronomy Review, 47, 677

\bibitem[{Daum {et~al.}(1997)Daum, Hermann, Hess, {et~al.}}]{daum:1997a}
Daum, A., Hermann, G., Hess, M., {et~al.} 1997, Astroparticle Physics, 8, 1

\bibitem[{{Dole} {et~al.}(2006){Dole}, {Lagache}, {Puget}, {Caputi},
  {Fern{\'a}ndez-Conde}, {Le Floc'h}, {Papovich}, {P{\'e}rez-Gonz{\'a}lez},
  {Rieke}, \& {Blaylock}}]{dole:2006a}
{Dole}, H., {Lagache}, G., {Puget}, J.-L., {et~al.} 2006, \aap, 451, 417

\bibitem[{{Dube} {et~al.}(1979){Dube}, {Wickes}, \& {Wilkinson}}]{dube:1979a}
{Dube}, R.~R., {Wickes}, W.~C., \& {Wilkinson}, D.~T. 1979, \apj, 232, 333

\bibitem[{Dwek \& Arendt(1998)}]{dwek:1998b}
Dwek, E. \& Arendt, R.~G. 1998, The Astrophysical Journal, 508, L9

\bibitem[{{Dwek} {et~al.}(2005{\natexlab{a}}){Dwek}, {Arendt}, \&
  {Krennrich}}]{dwek:2005c}
{Dwek}, E., {Arendt}, R.~G., \& {Krennrich}, F. 2005{\natexlab{a}}, \apj, 635,
  784

\bibitem[{{Dwek} \& {Krennrich}(2005)}]{dwek:2005a}
{Dwek}, E. \& {Krennrich}, F. 2005, \apj, 618, 657

\bibitem[{{Dwek} {et~al.}(2005{\natexlab{b}}){Dwek}, {Krennrich}, \&
  {Arendt}}]{dwek:2005b}
{Dwek}, E., {Krennrich}, F., \& {Arendt}, R.~G. 2005{\natexlab{b}}, \apj, 634,
  155

\bibitem[{{Dwek} \& {Slavin}(1994)}]{dwek:1994a}
{Dwek}, E. \& {Slavin}, J. 1994, \apj, 436, 696

\bibitem[{{Eadie} {et~al.}(1988){Eadie}, {Drijard}, {James}, {Roos}, \&
  {Sadoulet}}]{eadie:1998a}
{Eadie}, W.~T., {Drijard}, D., {James}, F.~E., {Roos}, M., \& {Sadoulet}, B.
  1988, Statistical Methods in Experimental Physics (North-Holland Publishing
  Company; Amsterdam, New-York, Oxford)

\bibitem[{{Edelstein} {et~al.}(2000){Edelstein}, {Bowyer}, \&
  {Lampton}}]{edelstein:2000a}
{Edelstein}, J., {Bowyer}, S., \& {Lampton}, M. 2000, \apj, 539, 187

\bibitem[{{Elbaz} {et~al.}(2002){Elbaz}, {Cesarsky}, {Chanial},
  {et~al.}}]{elbaz:2002a}
{Elbaz}, D., {Cesarsky}, C.~J., {Chanial}, P., {et~al.} 2002, Astronomy and
  Astrophysics, 384, 848

\bibitem[{{Fazio} {et~al.}(2004){Fazio}, {Ashby}, {Barmby}, {Hora}, {Huang},
  {Pahre}, {Wang}, {Willner}, {Arendt}, {Moseley}, {Brodwin}, {Eisenhardt},
  {Stern}, {Tollestrup}, \& {Wright}}]{fazio:2004a}
{Fazio}, G.~G., {Ashby}, M.~L.~N., {Barmby}, P., {et~al.} 2004, \apjs, 154, 39

\bibitem[{{Fazio} \& {Stecker}(1970)}]{fazio:1970a}
{Fazio}, G.~G. \& {Stecker}, F.~W. 1970, \nat, 226, 135

\bibitem[{{Finkbeiner} {et~al.}(2000){Finkbeiner}, Davis, \&
  Schlegel}]{finkbeiner:2000a}
{Finkbeiner}, D.~P., Davis, M., \& Schlegel, D.~J. 2000, The Astrophysical
  Journal, 544, 81

\bibitem[{{Finley} \& {The VERITAS Collaboration}(2001)}]{finley:2001a}
{Finley}, J.~P. \& {The VERITAS Collaboration}. 2001, in International Cosmic
  Ray Conference, 2827--+

\bibitem[{{Frayer} {et~al.}(2006){Frayer}, {Huynh}, {Chary}, {Dickinson},
  {Elbaz}, {Fadda}, {Surace}, {Teplitz}, {Yan}, \& {Mobasher}}]{frayer:2006a}
{Frayer}, D.~T., {Huynh}, M.~T., {Chary}, R., {et~al.} 2006, \apjl, 647, L9

\bibitem[{{Gaidos} {et~al.}(1996){Gaidos}, {Akerlof}, {Biller}, {Boyle},
  {Breslin}, {Buckley}, {Carter-Lewis}, {Catanese}, {Cawley}, {Fegan},
  {Finley}, {Hillas}, {Krennrich}, {Lamb}, {Lessard}, {McEnery}, {Mohanty},
  {Moriarty}, {Quinn}, {Rodgers}, {Rose}, {Samuelson}, {Schubnell},
  {Sembroski}, {Srinivasan}, {Weekes}, {Wilson}, \& {Zweerink}}]{gaidos:1996a}
{Gaidos}, J.~A., {Akerlof}, C.~W., {Biller}, S.~D., {et~al.} 1996, \nat, 383,
  319

\bibitem[{Gorjian {et~al.}(2000)Gorjian, Wright, \& Chary}]{gorjian:2000a}
Gorjian, V., Wright, E.~L., \& Chary, R.~R. 2000, The Astrophysical Journal,
  536, 550

\bibitem[{{Gould} \& {Schr{\'e}der}(1967)}]{gould:1967a}
{Gould}, R.~J. \& {Schr{\'e}der}, G.~P. 1967, Physical Review, 155, 1408

\bibitem[{{Guy} {et~al.}(2000){Guy}, {Renault}, {Aharonian}, {Rivoal}, \&
  {Tavernet}}]{guy:2000a}
{Guy}, J., {Renault}, C., {Aharonian}, F.~A., {Rivoal}, M., \& {Tavernet},
  J.-P. 2000, Astronomy and Astrophysics, 359, 419

\bibitem[{{Hauser} {et~al.}(1998){Hauser}, {Arendt}, {Kelsall}, {Dwek},
  {Odegard}, {et~al.}}]{hauser:1998a}
{Hauser}, M.~G., {Arendt}, R.~G., {Kelsall}, T., {et~al.} 1998, The
  Astrophysical Journal, 508, 25

\bibitem[{{Hauser} \& {Dwek}(2001)}]{hauser:2001a}
{Hauser}, M.~G. \& {Dwek}, E. 2001, Annual Review of Astronomy and
  Astrophysics, 39, 249

\bibitem[{Heitler(1960)}]{heitler:1960a}
Heitler, M. 1960, The Quantum Theorie of Radiation (Oxford, Clarendon)

\bibitem[{{Hinton}(2004)}]{Hinton2004:HESSStatus}
{Hinton}, J.~A. 2004, New Astronomy Review, 48, 331

\bibitem[{{Horan} \& {Weekes}(2004)}]{horan:2004a}
{Horan}, D. \& {Weekes}, T.~C. 2004, New Astronomy Review, 48, 527

\bibitem[{{Kashlinsky}(2005)}]{Kashlinsky2005:EBLReview}
{Kashlinsky}, A. 2005, \physrep, 409, 361

\bibitem[{{Kashlinsky} {et~al.}(2005){Kashlinsky}, {Arendt}, {Mather}, \&
  {Moseley}}]{kashlinsky:2005a}
{Kashlinsky}, A., {Arendt}, R.~G., {Mather}, J., \& {Moseley}, S.~H. 2005,
  \nat, 438, 45

\bibitem[{Kashlinsky {et~al.}(1996)Kashlinsky, Mather, Odenwald, \&
  Hauser}]{kashlinsky:1996a}
Kashlinsky, A., Mather, J., Odenwald, S., \& Hauser, M. 1996, The Astrophysical
  Journal, 470, 681

\bibitem[{Kashlinsky \& Odenwald(2000)}]{kashlinsky:2000a}
Kashlinsky, A. \& Odenwald, S. 2000, The Astrophysical Journal, 528, 74

\bibitem[{{Katarzy{\'n}ski} {et~al.}(2006){Katarzy{\'n}ski}, {Ghisellini},
  {Tavecchio}, {Gracia}, \& {Maraschi}}]{katarzynski:2006a}
{Katarzy{\'n}ski}, K., {Ghisellini}, G., {Tavecchio}, F., {Gracia}, J., \&
  {Maraschi}, L. 2006, \mnras, 368, L52

\bibitem[{{Kifune}(1999)}]{kifune:1999a}
{Kifune}, T. 1999, \apjl, 518, L21

\bibitem[{{Kneiske} {et~al.}(2004){Kneiske}, {Bretz}, {Mannheim}, \&
  {Hartmann}}]{kneiske:2004a}
{Kneiske}, T.~M., {Bretz}, T., {Mannheim}, K., \& {Hartmann}, D.~H. 2004, \aap,
  413, 807

\bibitem[{{Kneiske} {et~al.}(2002){Kneiske}, {Mannheim}, \&
  {Hartmann}}]{kneiske:2002a}
{Kneiske}, T.~M., {Mannheim}, K., \& {Hartmann}, D.~H. 2002, \aap, 386, 1

\bibitem[{{Krennrich} {et~al.}(2002){Krennrich}, {Bond}, {Bradbury}, {Buckley},
  {Carter-Lewis}, {Cui}, {de la Calle Perez}, {Fegan}, {Fegan}, {Finley},
  {Gaidos}, {Gibbs}, {Gillanders}, {Hall}, {Hillas}, {Holder}, {Horan},
  {Jordan}, {Kertzman}, {Kieda}, {Kildea}, {Knapp}, {Kosack}, {Lang},
  {LeBohec}, {Moriarty}, {M{\"u}ller}, {Ong}, {Pallassini}, {Petry}, {Quinn},
  {Reay}, {Reynolds}, {Rose}, {Sembroski}, {Sidwell}, {Stanton}, {Swordy},
  {Vassiliev}, {Wakely}, \& {Weekes}}]{krennrich:2002a}
{Krennrich}, F., {Bond}, I.~H., {Bradbury}, S.~M., {et~al.} 2002, \apjl, 575,
  L9

\bibitem[{Lagache \& Puget(2000)}]{lagache:2000a}
Lagache, G. \& Puget, J.-L. 2000, Astronomy and Astrophysics, 355, 17

\bibitem[{{Leinert} {et~al.}(1998){Leinert}, {Bowyer}, {Haikala}, {Hanner},
  {Hauser}, {Levasseur-Regourd}, {Mann}, {Mattila}, {Reach}, {Schlosser},
  {Staude}, {Toller}, {Weiland}, {Weinberg}, \& {Witt}}]{leinert:1998a}
{Leinert}, C., {Bowyer}, S., {Haikala}, L.~K., {et~al.} 1998, \aaps, 127, 1

\bibitem[{{Levenson} {et~al.}(2007){Levenson}, {Wright}, \&
  {Johnson}}]{levenson:2007a}
{Levenson}, L.~R., {Wright}, E.~L., \& {Johnson}, B.~D. 2007, ArXiv e-prints,
  704

\bibitem[{{Madau} \& {Pozzetti}(2000)}]{madau:2000a}
{Madau}, P. \& {Pozzetti}, L. 2000, Monthly Notices of the Royal Astronomical
  Society, 312, L9

\bibitem[{{Mapelli} {et~al.}(2004){Mapelli}, {Salvaterra}, \&
  {Ferrara}}]{mapelli:2004a}
{Mapelli}, M., {Salvaterra}, R., \& {Ferrara}, A. 2004, in Baryons in Dark
  Matter Halos, ed. R.~{Dettmar}, U.~{Klein}, \& P.~{Salucci}

\bibitem[{{Martin} {et~al.}(1991){Martin}, {Hurwitz}, \&
  {Bowyer}}]{martin:1991a}
{Martin}, C., {Hurwitz}, M., \& {Bowyer}, S. 1991, \apj, 379, 549

\bibitem[{{Matsumoto} {et~al.}(2005){Matsumoto}, {Matsuura}, {Murakami},
  {Tanaka}, {Freund}, {Lim}, {Cohen}, {Kawada}, \& {Noda}}]{matsumoto:2005a}
{Matsumoto}, T., {Matsuura}, S., {Murakami}, H., {et~al.} 2005, \apj, 626, 31

\bibitem[{{Mattila}(1990)}]{mattila:1990a}
{Mattila}, K. 1990, in IAU Symp. 139: The Galactic and Extragalactic Background
  Radiation, ed. S.~{Bowyer} \& C.~{Leinert}, 257--268

\bibitem[{{Matute} {et~al.}(2006){Matute}, {La Franca}, {Pozzi}, {Gruppioni},
  {Lari}, \& {Zamorani}}]{matute:2006a}
{Matute}, I., {La Franca}, F., {Pozzi}, F., {et~al.} 2006, \aap, 451, 443

\bibitem[{{Metcalfe} {et~al.}(2003){Metcalfe}, {Kneib}, {McBreen}, {Altieri},
  {Biviano}, {Delaney}, {Elbaz}, {Kessler}, {Leech}, {Okumura}, {Ott},
  {Perez-Martinez}, {Sanchez-Fernandez}, \& {Schulz}}]{metcalfe:2003a}
{Metcalfe}, L., {Kneib}, J.-P., {McBreen}, B., {et~al.} 2003, \aap, 407, 791

\bibitem[{{Papovich} {et~al.}(2004){Papovich}, {Dole}, {Egami}, {Le Floc'h},
  {P{\'e}rez-Gonz{\'a}lez}, {Alonso-Herrero}, {Bai}, {Beichman}, {Blaylock},
  {Engelbracht}, {Gordon}, {Hines}, {Misselt}, {Morrison}, {Mould},
  {Muzerolle}, {Neugebauer}, {Richards}, {Rieke}, {Rieke}, {Rigby}, {Su}, \&
  {Young}}]{papovich:2004a}
{Papovich}, C., {Dole}, H., {Egami}, E., {et~al.} 2004, \apjs, 154, 70

\bibitem[{Primack {et~al.}(1999)Primack, Bullock, Summerville, \&
  MacMinn}]{primack:1999a}
Primack, J., Bullock, J., Summerville, R., \& MacMinn, D. 1999, Astroparticle
  Physics, 11, 93

\bibitem[{{Primack} {et~al.}(2005){Primack}, {Bullock}, \&
  {Somerville}}]{primack:2005a}
{Primack}, J.~R., {Bullock}, J.~S., \& {Somerville}, R.~S. 2005, in AIP Conf.
  Proc. 745: High Energy Gamma-Ray Astronomy, ed. F.~A. {Aharonian}, H.~J.
  {V{\"o}lk}, \& D.~{Horns}, 23--33

\bibitem[{{Protheroe} \& {Meyer}(2000)}]{protheroe:2000a}
{Protheroe}, R.~J. \& {Meyer}, H. 2000, Physics Letters B, 493, 1

\bibitem[{{Punch} {et~al.}(1992){Punch}, {Akerlof}, {Cawley},
  {et~al.}}]{punch:1992a}
{Punch}, M., {Akerlof}, C.~W., {Cawley}, M.~F., {et~al.} 1992, Nature, 358, 477

\bibitem[{{Salvaterra} \& {Ferrara}(2003)}]{salvaterra:2003a}
{Salvaterra}, R. \& {Ferrara}, A. 2003, Monthly Notice of the Royal
  Astronomical Society, 339, 973

\bibitem[{{Salvaterra} \& {Ferrara}(2006)}]{salvaterra:2006a}
{Salvaterra}, R. \& {Ferrara}, A. 2006, \mnras, 367, L11

\bibitem[{{Schroedter} {et~al.}(2005){Schroedter}, {Badran}, {Buckley},
  {Bussons Gordo}, {Carter-Lewis}, {Duke}, {Fegan}, {Fegan}, {Finley},
  {Gillanders}, {Grube}, {Horan}, {Kenny}, {Kertzman}, {Kosack}, {Krennrich},
  {Kieda}, {Kildea}, {Lang}, {Lee}, {Moriarty}, {Quinn}, {Quinn},
  {Power-Mooney}, {Sembroski}, {Wakely}, {Vassiliev}, {Weekes}, \&
  {Zweerink}}]{schroedter:2005a}
{Schroedter}, M., {Badran}, H.~M., {Buckley}, J.~H., {et~al.} 2005, \apj, 634,
  947

\bibitem[{{Stecker} {et~al.}(1996){Stecker}, {de Jager}, \&
  {Salamon}}]{stecker:1996a}
{Stecker}, F.~W., {de Jager}, O.~C., \& {Salamon}, M.~H. 1996, \apjl, 473, L75+

\bibitem[{{Stecker} \& {Glashow}(2001)}]{stecker:2001a}
{Stecker}, F.~W. \& {Glashow}, S.~L. 2001, Astroparticle Physics, 16, 97

\bibitem[{{Stecker} {et~al.}(2006){Stecker}, {Malkan}, \&
  {Scully}}]{stecker:2006a}
{Stecker}, F.~W., {Malkan}, M.~A., \& {Scully}, S.~T. 2006, \apj, 648, 774

\bibitem[{{Toller}(1983)}]{toller:1983a}
{Toller}, G.~N. 1983, \apjl, 266, L79

\bibitem[{Wright \& Reese(2000)}]{wright:2000a}
Wright, E.~L. \& Reese, E.~D. 2000, The Astrophysical Journal, 545, 43

\end{thebibliography}

\end{document}